\documentclass[fleqn,usenatbib]{mnras}

\usepackage{newtxtext,newtxmath}

\usepackage[T1]{fontenc}

\DeclareRobustCommand{\VAN}[3]{#2}
\let\VANthebibliography\thebibliography
\def\thebibliography{\DeclareRobustCommand{\VAN}[3]{##3}\VANthebibliography}

\usepackage{graphicx}
\usepackage{amsmath} 
\usepackage{tablefootnote}
\usepackage{longtable}
\usepackage{amsfonts}
\usepackage{soul}
\usepackage{hyperref}
\usepackage{enumitem}
\usepackage[hyphens]{xurl}
\usepackage{orcidlink}

\title[Survey of $z>3$ Ly$\alpha$ Emitters in the EGS]{Deeper than DEEP: A Spectroscopic Survey of \boldmath$z>3$ Lyman-\boldmath$\alpha$ Emitters in the Extended Groth Strip\thanks{The data presented herein were obtained at the W. M. Keck Observatory, which is operated as a scientific partnership among the California Institute of Technology, the University of California and the National Aeronautics and Space Administration. The Observatory was made possible by the generous financial support of the W. M. Keck Foundation.} }
    
\author[Urbano Stawinski et al.]{Stephanie M. Urbano Stawinski\textsuperscript{\orcidlink{0000-0001-8169-7249}},$^{1}$\textsuperscript{\thanks{E-mail: ststawinski@gmail.com}}
M. C. Cooper\textsuperscript{\orcidlink{0000-0003-1371-6019}},$^{1}$
Steven L. Finkelstein\textsuperscript{\orcidlink{0000-0001-8519-1130}},$^{2}$
Intae Jung\textsuperscript{\orcidlink{0000-0003-1187-4240}},$^{3}$
\newauthor
Pablo G. P\'erez-Gonz\'alez\textsuperscript{\orcidlink{0000-0003-4528-5639}},$^{4}$
Micaela B. Bagley\textsuperscript{\orcidlink{0000-0002-9921-9218}},$^{2}$
Caitlin M. Casey\textsuperscript{\orcidlink{0000-0002-0930-6466}},$^{2}$
Olivia R. Cooper\textsuperscript{\orcidlink{0000-0003-3881-1397}},$^{2}$
\newauthor
Nimish P. Hathi\textsuperscript{\orcidlink{0000-0001-6145-5090}},$^{3}$
Benne W. Holwerda\textsuperscript{\orcidlink{0000-0002-4884-6756}},$^{5}$
Anton M. Koekemoer\textsuperscript{\orcidlink{0000-0002-6610-2048}},$^{3}$
Jeyhan S. Kartaltepe\textsuperscript{\orcidlink{0000-0001-9187-3605}},$^{6}$
\newauthor
Vital Fern\'{a}ndez\textsuperscript{\orcidlink{0000-0003-0531-5450}},$^{7}$
Rebecca L. Larson\textsuperscript{\orcidlink{0000-0003-2366-8858}},$^{6}$
Ray A. Lucas\textsuperscript{\orcidlink{0000-0003-1581-7825}},$^{3}$
L. Y. Aaron Yung\textsuperscript{\orcidlink{0000-0003-3466-035X}}$^{8}$ \\ \\
$^{1}${Department of Physics \& Astronomy, University of California, Irvine, 4129 Reines Hall, Irvine, CA 92697, USA}\\
$^{2}${Department of Astronomy, The University of Texas at Austin, Austin, TX 78712 USA} \\
$^{3}${Space Telescope Science Institute, 3700 San Martin Dr., Baltimore, MD 21218, USA} \\
$^{4}${Centro de Astrobiolog\'{\i}a (CAB), CSIC-INTA, Ctra. de Ajalvir km 4, Torrej\'on de Ardoz, E-28850, Madrid, Spain}\\
$^{5}${Physics \& Astronomy Department, University of Louisville, Louisville, KY, 40292 USA} \\
$^{6}${School of Physics and Astronomy, Rochester Institute of Technology, 84 Lomb Memorial Drive, Rochester, NY 14623, USA} \\
$^{7}${Instituto de Investigaci\'{o}n Multidisciplinar en Ciencia y Tecnolog\'{i}a, Universidad de La Serena, Raul Bitr\'{a}n 1305, La Serena 2204000, Chile} \\ 
$^{8}${Astrophysics Science Division, NASA Goddard Space Flight Center, 8800 Greenbelt Rd, Greenbelt, MD 20771, USA}\\}

\pubyear{2023}

\begin{document}
\label{firstpage}
\pagerange{\pageref{firstpage}--\pageref{lastpage}}
\maketitle

\begin{abstract}
    We present a spectroscopic survey of Ly$\alpha$ emitters in the Extended Groth Strip (EGS) field, targeting the regime near the Epoch of Reionization. Using Keck/DEIMOS, we observed 947 high-$z$ candidates with photometric redshifts from $3 < z_\text{phot} < 7$ and down to an $H$-band ({\it{HST}}/WFC3 F160W) magnitude limit of $< 27.5$. Observations were taken over the course of 8 nights, with integration times ranging from 4 to 7.8 hours. Our survey secured 137 unique redshifts, 126 of which are Ly$\alpha$ emitters at $2.8 < z < 6.3$ with a mean redshift of $\overline{z} = 4.3$. We provide a comprehensive redshift catalog for our targets, as well as the reduced one- and two- dimensional spectra for each object. These observations will provide an important auxiliary dataset for the {\textit{JWST}} Directors Discretionary Early Release Science (DD-ERS) program the Cosmic Evolution Early Release Science Survey (CEERS), which recently completed near- and mid-IR imaging and spectroscopy of galaxies in the EGS field.
\end{abstract}


\section{Introduction}
Deep surveys in widely studied extragalactic fields are pivotal in characterizing galaxy evolution across cosmic time. The Extended Groth Strip (EGS) is one of the leading extragalactic fields on the sky, renowned for a balance of area and depth with observations extending from X-ray to radio wavelengths \citep[e.g.][]{Davis_2007, Ivison_2007, Willner_2012, Nandra_2015}. The EGS field is centered at $\alpha = 14^\text{h}19^\text{m}00^\text{s}$ and $\delta = +52^\circ48^\text{m}00^\text{s}$ with the bulk of deep imaging observations covering a central region of $800$ square arcminutes. Its relevance in extragalactic astronomy is due in part to major surveys using a variety of instruments, including the {\it{Hubble Space Telescope (HST)}} through both the All-wavelength Extended Groth strip International Survey (AEGIS, \citealt{Davis_2007}) and the Cosmic Assembly Near-infrared Deep Extragalactic Legacy Survey (CANDELS, \citealt{Grogin_2011,Koekemoer_2011}). Now with the launch of {\it{JWST}} \citep{Gardner_2023,Gardner_2006}, the EGS has further cemented its status as a legacy field, due predominately to the Cosmic Evolution Early Release Science (CEERS) Survey (ERS 1345, PI: S Finkelstein\footnote{CEERS data can be publicly accessed in MAST: \href{https://doi.org/10.17909/z7p0-8481}{10.17909/z7p0-8481}}), a Director's Discretionary Early Release Science (DD-ERS) program that has conducted both imaging and spectroscopy with {\it{JWST}} in the EGS (\citealt{Finkelstein_2023,Finkelstein_2022_maisie}; Finkelstein et al.~in prep).

The significant amount of existing observations and telescope time dedicated to the EGS makes supplemental spectroscopic observations increasingly powerful. Spectroscopic observations have routinely been used for confirmation or readjustment of photometric redshifts, ultimately improving the reliability of constraints derived from photometric spectral energy distribution (SED) fitting. Spectroscopy is also critical for obtaining certain spectral properties and dynamical measurements (such as emission and absorption line strengths, velocity offsets, and velocity widths). For these reasons, spectroscopic data drastically improve the implied constraints from photometry alone. 

\begin{figure}
	\centering
\includegraphics[width=\linewidth]{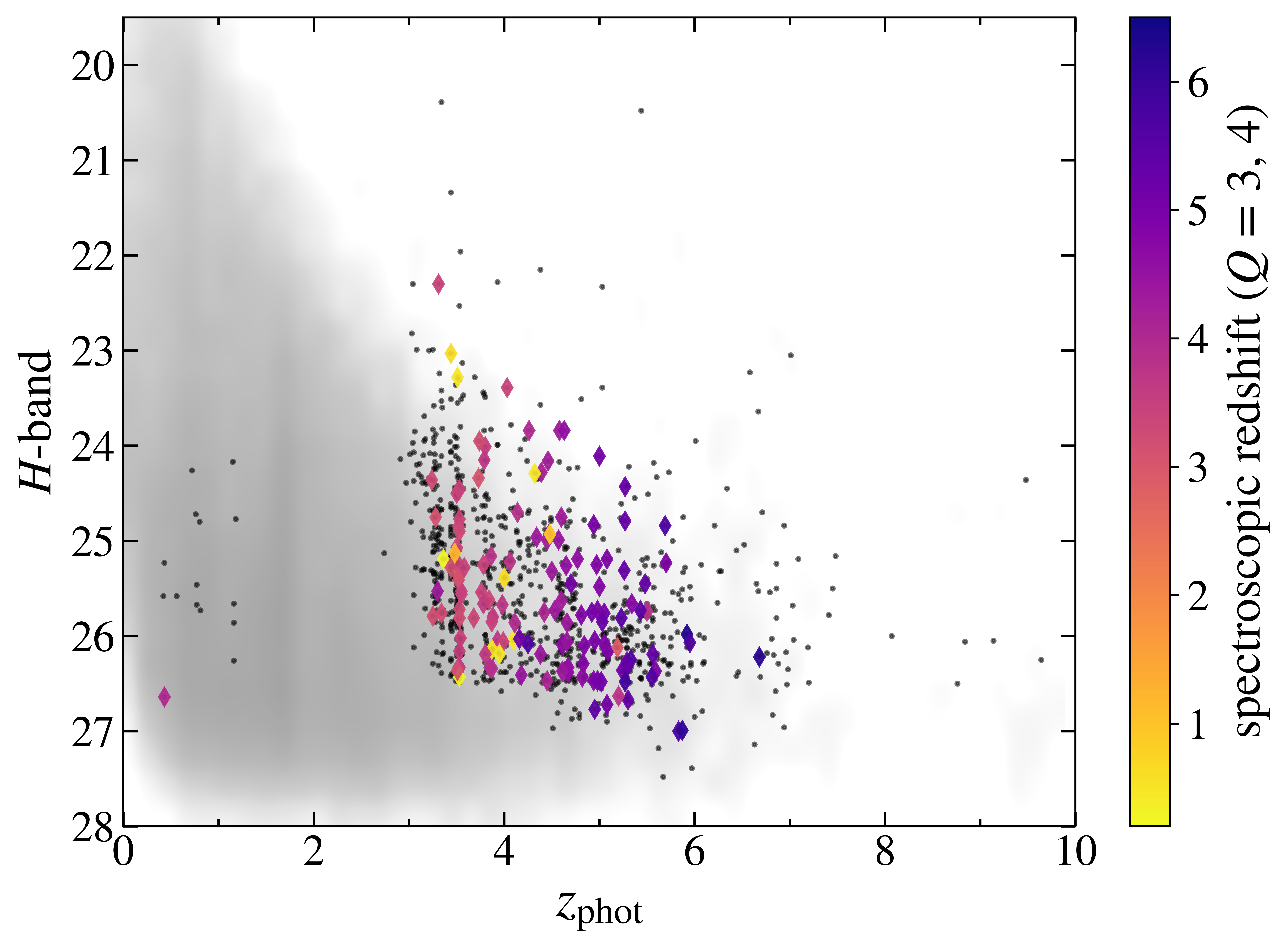}
		\caption{Apparent $H$-band ({\it{HST}}/WFC3 F160W) magnitude versus Photometric redshift ($z_\text{phot}$) for the target sample from \citet{Finkelstein_2022}. Secure redshifts from this work (i.e.~having a quality flag $Q = 3,4$, see \S\ref{sec:obs}) are shown as diamonds color-coded according to their measured spectroscopic redshift. Black dots show sources that were targeted but did not yield a secure redshift. The grey density map illustrates the distribution of sources from the complete \citet{Stefanon_2017} catalog.
\label{fig:Hmag}}
\end{figure}

In June 2022, CEERS began imaging in the EGS using the Near Infrared Camera (NIRCam, \citealt{Rieke_2003,Beichman_2012}) and the Mid-Infrared Instrument \citep[MIRI,][]{Rieke_2015,Wright_2015} and continued to obtain additional photometric imaging in December 2022. The CEERS collaboration has since published the first data release of NIRCam observations \citep{Bagley_2023} and MIRI imaging \citep{Papovich_2023,Yang_2023}. With the influx of photometry in the EGS field using {\textit{JWST}} in the first year of operation, spectroscopic catalogs at high-$z$ become particularly useful for improving inferred galaxy properties and informing future science. 

Foundational spectroscopic surveys in the EGS, such as the DEEP2 and DEEP3 surveys (\citealt{Newman_2013, Cooper_2012_deep3, Zhou_2019}, see also \citealt{Weiner_2005}) have provided extensive spectroscopy within the EGS, greatly improving the inferred constraints on galaxy properties at low and intermediate redshifts. While the DEEP2 and DEEP3 surveys provide highly uniform spectra and an extremely high sampling density of secure redshifts at $z \lesssim 1$, the EGS has trailed other extragalactic fields, such as COSMOS, GOODS-N, and GOODS-S, with respect to spectroscopic coverage at higher $z$ \citep[e.g.][]{Reddy_2006, Vanzella_2008, Vanzella_2009, LeFevre_2015, Silverman_2015, Pentericci_2018b, Pentericci_2018a, McLure_2018, Hasinger_2018}. Over the past decade, however, the 3D-HST survey \citep{Momcheva_2016,Brammer_2012} as well as the MOSFIRE Deep Evolution Field survey (MOSDEF, \citealt{Kriek_2015}) both began to push spectroscopic studies to higher $z$ in the EGS field. 3D-HST used {\it HST} WFC3-IR/G141 grism spectroscopy to measure $\sim 3000$ secure grism redshifts, including $\sim 500$ galaxies at $2 < z < 3$ and another $26$ at $3 < z < 3.5$. In addition, MOSDEF targeted $\sim 1500$ galaxies at $1.37 < z < 3.80$ from the EGS, GOODS-N, and COSMOS fields. Despite these more recent near-IR spectroscopic campaigns along with smaller efforts to study very high-$z$ sources \citep[e.g.][]{Oesch_2015, Zitrin_2015, RB_2016, Tilvi_2020,   Larson_2023, Jung_2022}, the EGS is still lacking in spectral coverage for galaxies at $z > 4$, an important epoch with respect to the recently-completed CEERS {\textit{JWST}} photometric observations. CEERS spectroscopic observations using {\it JWST} NIRSpec and NIRcam are poised to greatly increase the publicly-available spectroscopy of high-$z$ galaxies in the EGS \citep[e.g.][]{Shapley_2023, Reddy_2023, Isobe_2023}, yielding redshifts for hundreds of sources at a range of $z$, including at $z \sim 8-10$ \citep{Jung_2023,Fujimoto_2023,ArrabalHaro2023,Larson_2022,Harikane_2024,Tang_2023}.

\begin{figure}
	\centering
    \includegraphics[width=\linewidth]{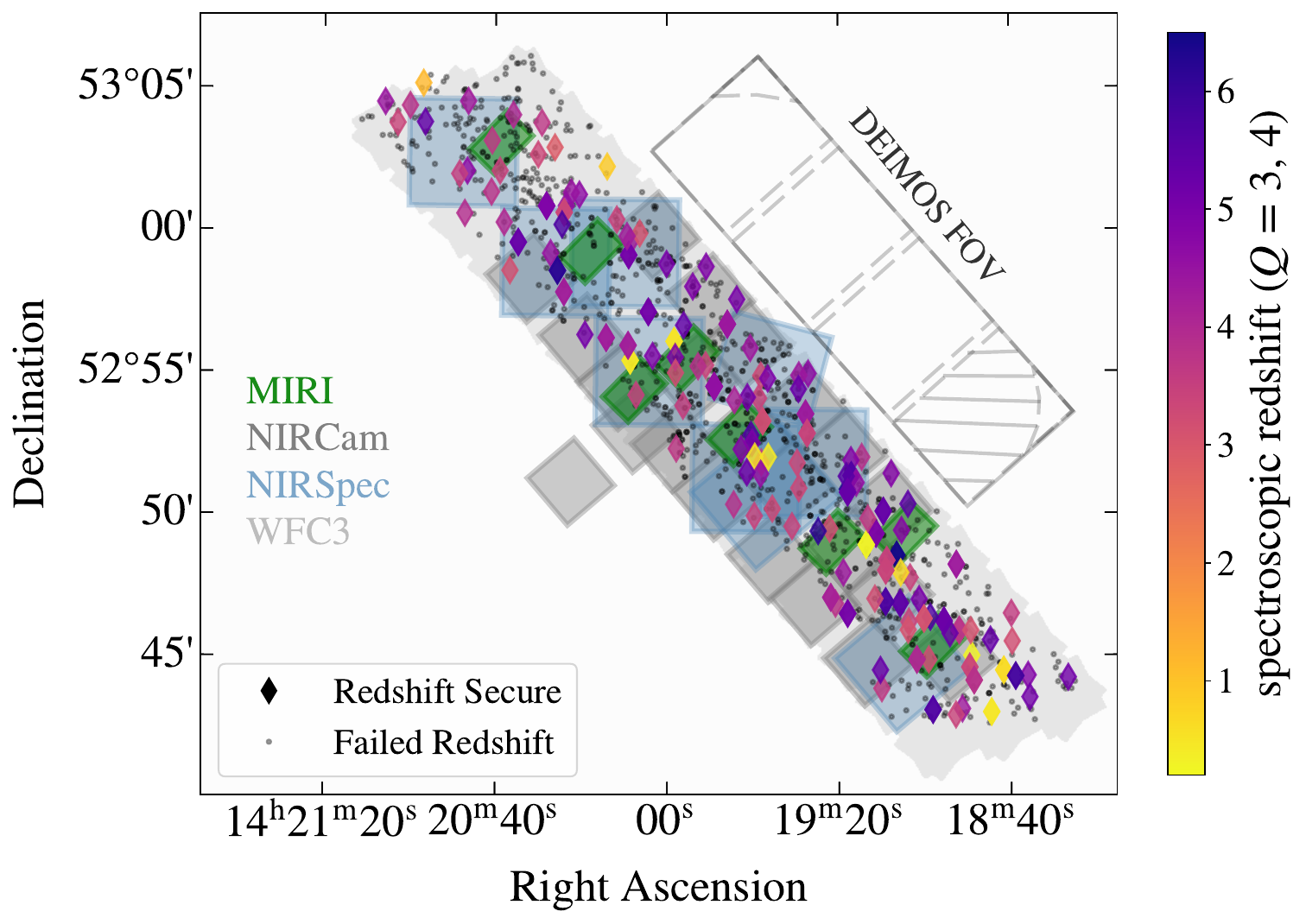}
		\caption{Distribution of Keck/DEIMOS targets on the sky. Sources with secure redshifts ($Q = 3, 4$; see \S\ref{sec:obs}) are displayed as diamonds,  color-coded by the measured spectroscopic redshift, with all other targets ($Q = -2, 1, 2$) plotted as grey dots. The CEERS pointings for MIRI, NIRCam, and NIRSpec are shown as green, dark grey, and blue rectangles, respectively. The shaded light grey region displays the CANDELS {\it{HST}}/WFC3 footprint in the EGS. In addition, we overplot the Keck/DEIMOS FOV and designate the portion of slitmask real estate affected by the inoperable CCD5 with hash marks (see \S\ref{sec:obs}). 
\label{fig:obs_map}}
\end{figure}

To further supplement the recent deep, near- and mid-IR imaging data in the EGS from {\it{JWST}}, we undertook spectroscopic observations of intermediate- and high-$z$ sources using the DEep Imaging Multi-Object Spectrograph (DEIMOS; \citealt{Faber_2003}) on the KECK II telescope. We present spectroscopic observations of $947$ targets with $137$ unique spectroscopically confirmed redshifts. The majority (126) of these objects are Ly$\alpha$ emitters at $2.8 < z < 6.5$, increasing the spectroscopic coverage of high-$z$ galaxies in the EGS. In Sections \ref{sec:selection} and \ref{sec:obs}, we describe our target selection and observations for the survey, respectively. We present the Keck/DEIMOS redshift catalog in Section \ref{sec:data}, along with subsequent analysis. Finally, in Section \ref{sec:conc}, we conclude with a discussion of the potential use of our survey and our recent collaborations with ongoing, ground-based high-$z$ surveys in the EGS.

\section{Target Selection and Slit Mask Design} \label{sec:selection}

\begin{table*}
	\caption{Slitmask Observation Information
	}
	\setlength{\tabcolsep}{15pt}
	\label{table:obs}
	\centering
    \begin{tabular}{cccccccc}
	    \hline \hline
	    Slit Mask & Observation & $\alpha$ (J2000)\textsuperscript{a} & $\delta$ (J2000)\textsuperscript{b} & PA\textsuperscript{c} & Exposure & $N_{\rm obj}$\textsuperscript{d} & $N_z$\textsuperscript{e}  \\
	    Number & Date (UT) & & & (deg) & Time (s) & & \\
	    \hline
1 &  2020-06-10  &  14:19:05.40  &  +52:49:27.0  &  220  &  20850  &  142 (39)  &  24  \\
2 &  2020-06-11  &  14:19:16.40  &  +52:49:31.5  &  222  &  28151  &  142 (37)  &  30  \\
3 &  2020-06-14  &  14:20:30.12  &  +52:59:14.3  &  44  &  28601  &  145 (37)  &  20  \\
4 &  2020-06-12  &  14:20:05.93  &  +52:59:19.5  &  220  &  13600  &  142 (37)  &  16  \\
21 &  2021-06-04  &  14:19:05.59  &  +52:49:28.0  &  220  &  14400  &  141 (35)  &  24  \\
22 &  2021-06-03  &  14:19:16.43  &  +52:49:28.2  &  222  &  14400  &  150 (37)  &  29  \\
23 &  2021-06-12  &  14:20:29.69  &  +52:59:15.4  &  44  &  14400  &  140 (34)  &  21  \\
24 &  2021-06-11  &  14:20:05.73  &  +52:59:21.6  &  220  &  14400  &  145 (34)  &  19  \\
	\hline
	\multicolumn{8}{l}{\textsuperscript{a}\footnotesize{Right ascension of slit mask in HH:MM:SS}}\\
	\multicolumn{8}{l}{\textsuperscript{b}\footnotesize{Declination of slit mask center in DD:MM:SS}}\\
	\multicolumn{8}{l}{\textsuperscript{c}\footnotesize{Position angle of slit mask in degrees (E of N)}}\\
	\multicolumn{8}{l}{\textsuperscript{d}\footnotesize{Number of total targets (Number of targets with spectra on DEIMOS CCD5)}}\\
	\multicolumn{8}{l}{\textsuperscript{e}\footnotesize{Number of secure redshifts ($Q =$ 3, 4)}}\\
	\end{tabular}
\end{table*}

As its name suggests, the EGS field spans an extended, narrow area on the sky. To efficiently explore Ly$\alpha$ emission from $z = 3-6$ over the entirety of the field as probed by the CANDELS {\it HST} imaging, we required an optical spectrograph with a similarly broad field-of-view (FOV) and capable of a high level of multiplexing. The Keck/DEIMOS spectrograph is particularly well-suited due to its large FOV that matches the shape of the CANDELS footprint in the EGS as well its ability to observe $\gtrsim 140$ targets simultaneously. 

Spectroscopic targets were selected from the photometric catalog of \citet{Finkelstein_2022}, based upon the CANDELS {\it{HST}} and {\it{Spitzer}} IRAC observations in the EGS. In this target catalog, two separate photometric redshifts for each source were dervied from \texttt{EAZY} (\citealt{Brammer_2008}): one including all CANDELS photometric measurements and one excluding IRAC photometry to avoid issues with poor deblending for higher redshift candidates. From this catalog, our primary targets were selected based on two selection criteria. The first is that targets have $3 < z_{\rm phot} < 7$ corresponding to the peak of the redshift probability distribution derived from \texttt{EAZY} from \textit{either} method, with or without IRAC photometry. The second criterion is for targets to have an $H$-band magnitude of $H < 27.5$, with priority given to brighter sources. We adopted this magnitude limit to avoid potentially spurious sources at the detection limit of the existing {\it HST} imaging, while the photometric redshift limits were chosen to match the lowest and highest $z$ at which Lyman-$\alpha$ would be detected given our instrument configuration (see \S\ref{sec:obs}). In summary, the original parent catalog included 3550 sources in the EGS. After selecting based on $z_{\rm phot}$, we trimmed the catalog to 2553 sources, with the magnitude cut further limiting the sample to 2518 primary targets used to populate slit masks, 911 of which ended up on at least one slit mask. To maximize the observations, slit masks were then filled with \textit{any} additional sources with $z_{\rm phot} > 3$ and $H < 27.5$ as well as (preferentially brighter) fillers at $z_{\rm phot} < 3$ without a secure spectroscopic redshift in the literature. In total, the target population included $947$ unique sources, with the vast majority (98\%) selected to be at $z_{\rm phot} > 3$ from the \citet{Finkelstein_2022} photo-$z$ catalog (including $96\%$ of targets at $3 < z_{\rm phot} < 7$). Figure~\ref{fig:Hmag} shows the distribution of our targeted sources, as well as filler targets, as a function of $z_{\rm phot}$ and $H$-band (F160W) magnitude, highlighting those that yielded a secure spectroscopic redshift (see \S\ref{sec:data}). 

We tiled the EGS with a total of $8$ slitmasks, located at 4 overlapping positions along the strip, such that sources had at least two opportunities to be placed on a mask. Table~\ref{table:obs} summarizes the position and number of targets for each slitmask, along with the date of observation and total exposure time. Across all masks, slit widths were fixed to $1^{\prime\prime}$, with slit gaps measuring $0{\farcs}5$, so as to optimize the number of potential targets observed on a given mask. Slit lengths were allowed to vary, above a minimum slit length of $4^{\prime\prime}$, such that slits were sufficiently long to avoid any significant loss in redshift success (see results from DEEP2/DEEP3, \citealt{Newman_2013,Cooper_2011,Cooper_2012_deep3,Zhou_2019}). Each slitmask includes roughly 140-150 sources per mask, with the final targeted sample including $947$ unique sources down to the magnitude limit of $H < 27.5$. Across the $8$ masks, $173$ sources are targeted more than once, with $22$ of these repeat targets placed on more than two masks. We describe how we handle repeat redshift measurements later in \S\ref{subsec:multiObs}. Figure~\ref{fig:obs_map} shows the distribution of targets with respect to the CANDELS {\it HST}/WFC3 imaging footprint as well as the CEERS {\it JWST} NIRCam, MIRI, and NIRSpec pointings. Along the southeast edge of the strip, overlap with the CEERS NIRCam fields is sub-optimal due to a lack of bright guide stars; though, the Keck/DEIMOS spectroscopy covers the central portion of the -- \emph{at that time} -- planned CEERS observations. 

\section{DEIMOS Observations and Reductions} \label{sec:obs}

As detailed in Table~\ref{table:obs}, spectroscopic observations were completed during June 2020 and 2021, prior to the launch of {\it JWST} in late 2021. With Keck/DEIMOS, we used the 600 lines mm$^{-1}$ grating blazed at 7500~\AA \ and tilted to a central wavelength of 7200~\AA, with the GG455 order-blocking filter employed. This spectroscopic setup provides an approximate spectral coverage of $\sim$ 4500–9900~\AA, depending on the slit placement on the particular mask. The spectral resolution (FWHM) for the 600g grating on DEIMOS is $\sim 3.5$~\AA\ \citep{Weiner_2006}, with a dispersion of $0.65$~\AA\ per pixel. 

Each individual exposure was typically $\sim$ 1800 sec in length, with a minimum of $7$ exposures per mask and no dithering applied between exposures. The total integration times achieved for each mask are listed in Table~\ref{table:obs}, ranging from $\lesssim 4$ hours to as much as $\sim 7.8$ hours. Calibrations for each mask included three internal quartz lamp flat-field frames and an arc lamp spectrum (using Kr, Ar, Ne, and Xe lamps). During observations, the DEIMOS flexure compensation system was utilized to ensure flexure frame-to-frame throughout the night (for both calibration and science images) differed by $\lesssim$ $\pm 0.25$ pixels. 

\begin{figure*}
	\centering
	\includegraphics[width=0.33\linewidth]{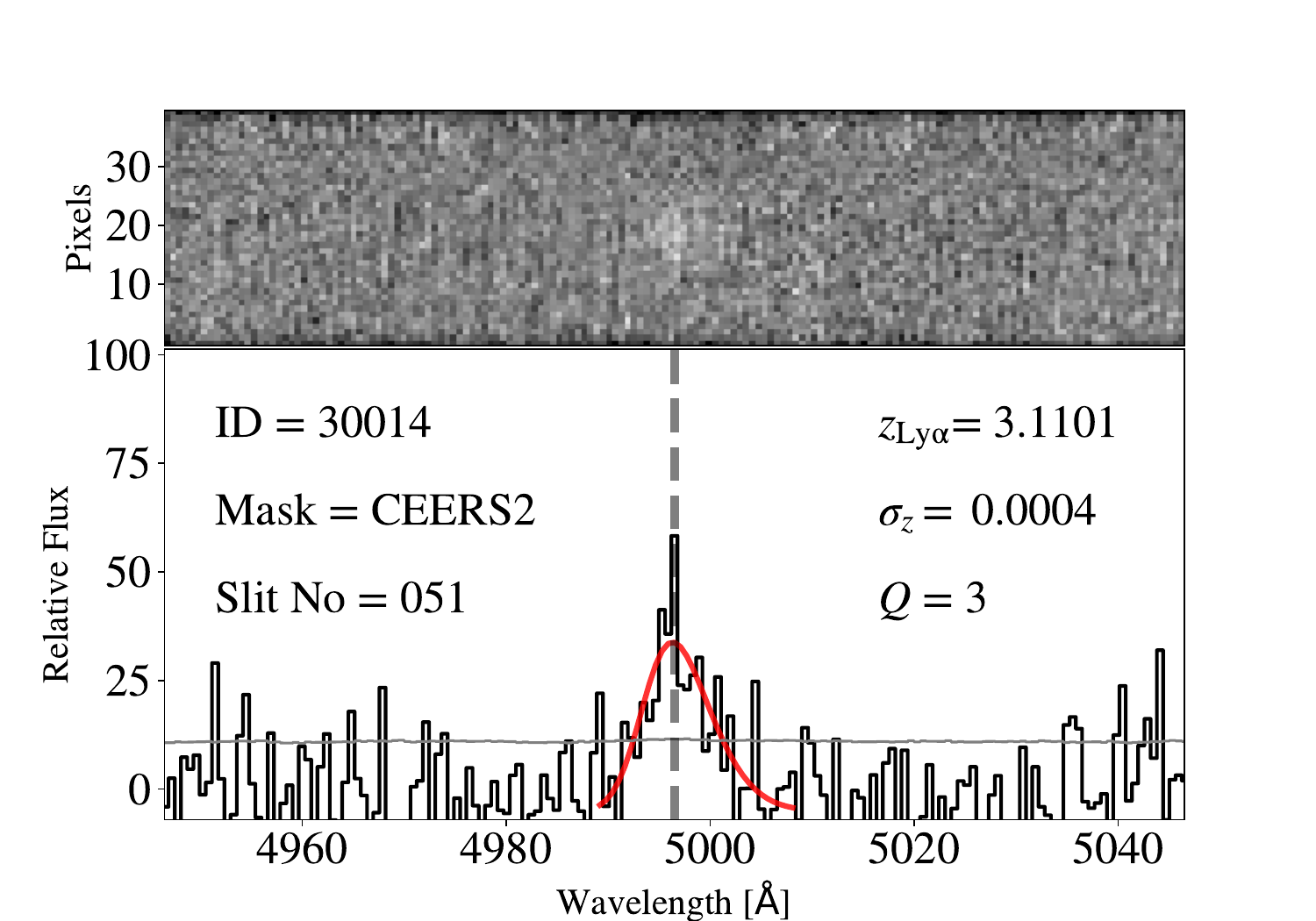}\includegraphics[width=0.33\linewidth]{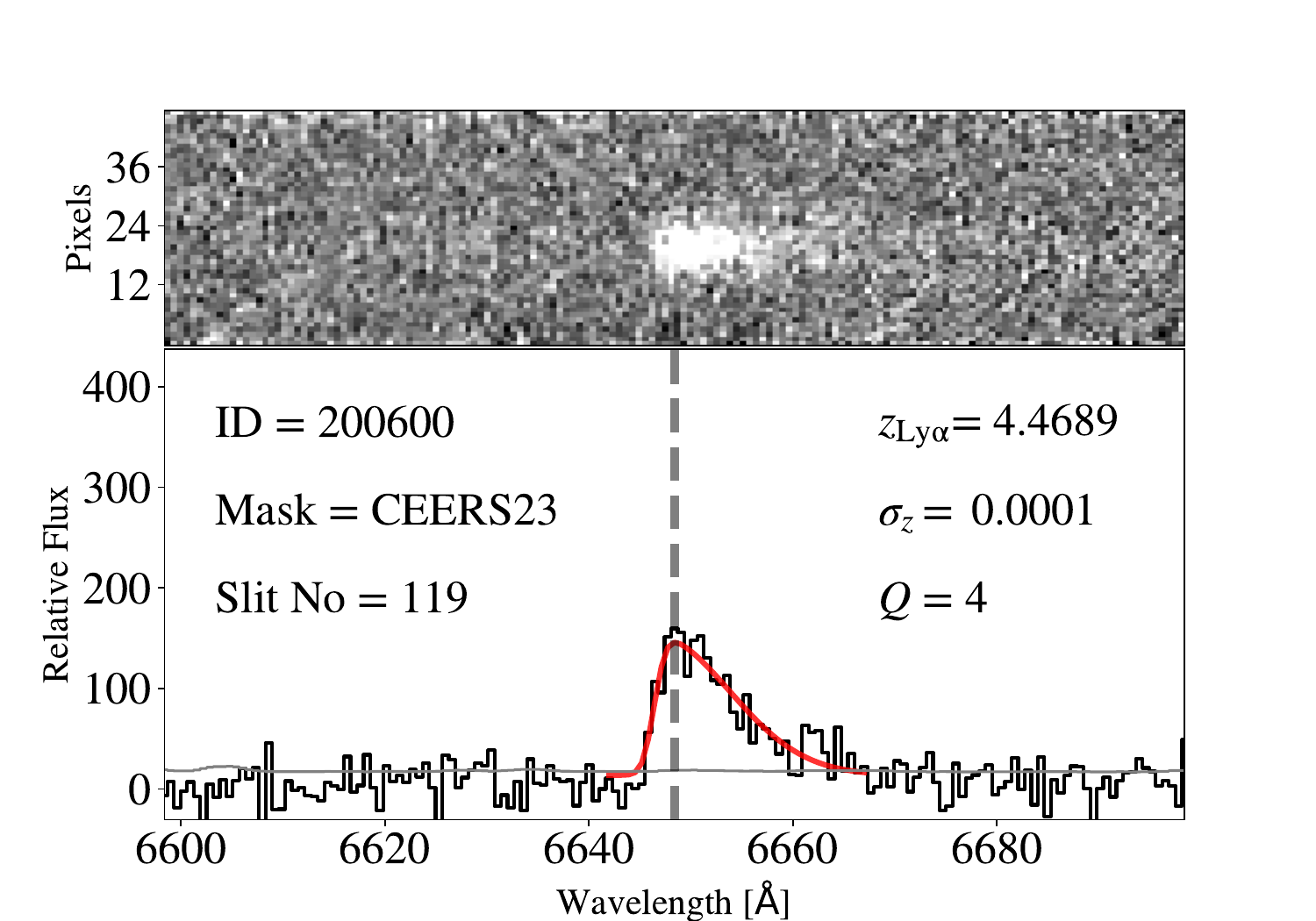}
    \includegraphics[width=0.33\linewidth]{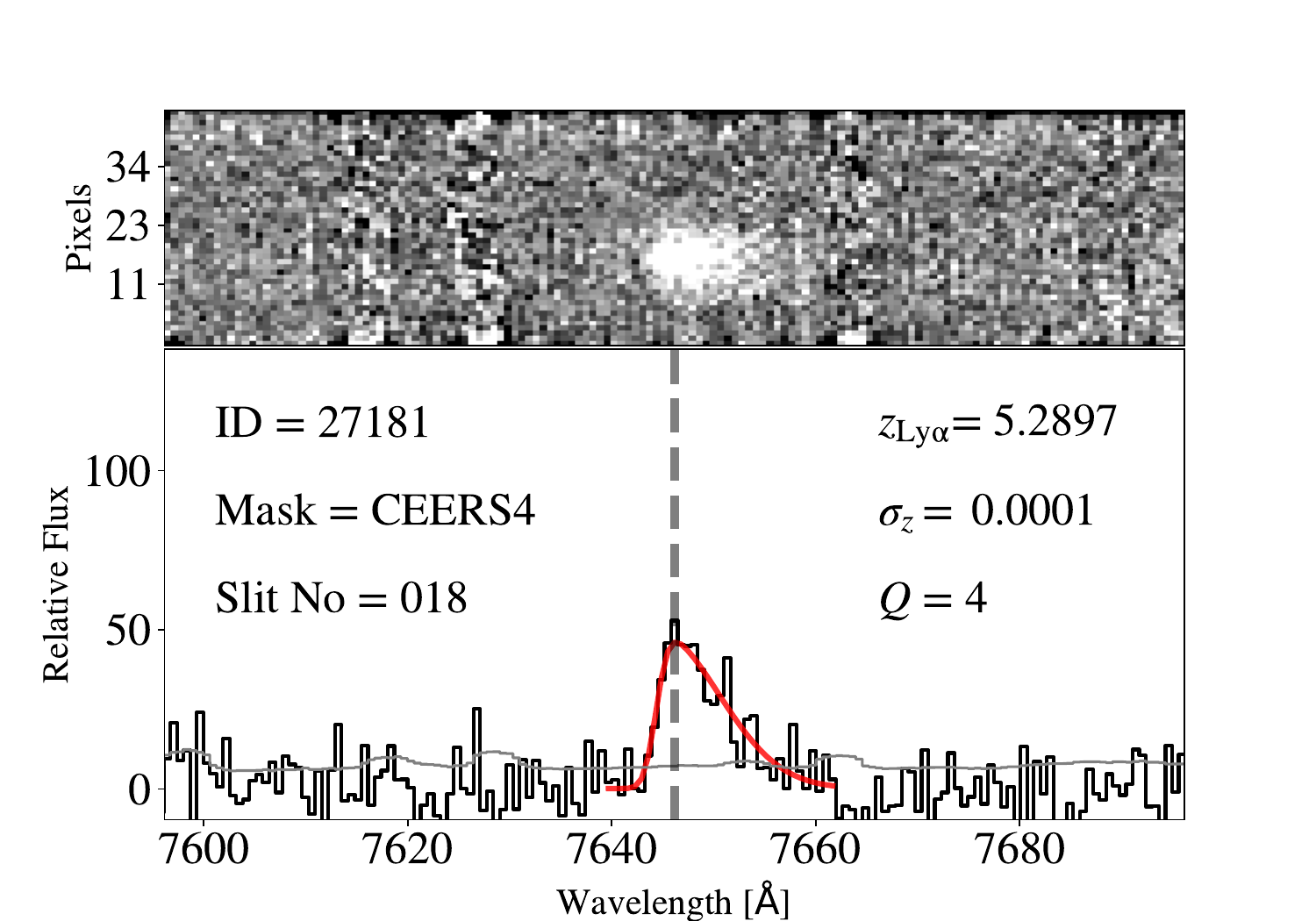}
		\caption{Examples of the 1D (bottom-panels) and 2D (top-panels) spectra for three confirmed Ly$\alpha$ emitters and the asymmetric Gaussian fit (shown in red) used to determine the redshift. The vertical dashed line shows the location of the peak Ly$\alpha$ emission at which we measure a redshift. The 1D variance array is shown in light grey. 
\label{fig:temp_ex}}
\end{figure*}

\begin{table*}
	\caption{Redshift Catalog}
	\setlength{\tabcolsep}{5pt}
	\label{table: cat}
	\centering
    \begin{tabular}{cccccccccccc}
	    \hline \hline
	    Object ID\textsuperscript{a} & Object ID\textsuperscript{b}  & Mask\textsuperscript{c} & Slit\textsuperscript{d} & MJD\textsuperscript{e} & $\alpha$ (J2000)\textsuperscript{f} & $\delta$ (J2000)\textsuperscript{g}  & $z_\text{obs}$\textsuperscript{h} & $z_\text{helio}$\textsuperscript{i} & $Q$ flag\textsuperscript{j} & F160W\textsuperscript{k} & $z_\text{phot}$\textsuperscript{l} \\
	     & Stefanon+17 &  &  & &  &  &  &  &  &  ({\it{HST}}/WFC3) & Stefanon+17 \\
	    \hline
11885 &  33390  &  CEERS22  &  5  &  59368.3815  &  214.791847  &  52.730646  &  3.6289  &  3.6289  &  3  &  25.80  &  3.317  \\
25744 &  37899  &  CEERS3  &  55  &  59014.2542  &  215.037178  &  52.984911  &  5.0789  &  5.0789  &  3  &  26.72  &  4.598  \\
26906 & 19503 & CEERS24 & 12 & 59376.4073 & 215.093025 & 53.021548 & 4.4730 & 4.4730 & 4 & 26.06 & 4.430 \\
30677 &  17738  &  CEERS2  &  56  &  59011.2535  &  214.702437  &  52.735304  &  3.8200  &  3.8200  &  4  &  26.33  &  3.750  \\
200600 &  9909  &  CEERS23  &  119  &  59377.2499  &  215.113233  &  52.985967  &  4.4689  &  4.4689  &  4  &  24.99  &  4.455  \\
	    \hline
	\multicolumn{12}{l}{\textsuperscript{a}\footnotesize{Object identification number from this work}}\\
    \multicolumn{12}{l}{\textsuperscript{b}\footnotesize{Object identification number from \citet{Stefanon_2017}}}\\
    \multicolumn{12}{l}
    {\textsuperscript{c}\footnotesize{Name of slit mask on which the object was observed}}\\
    \multicolumn{12}{l}
    {\textsuperscript{d}\footnotesize{Slit number on the corresponding mask}}\\
    \multicolumn{12}{l}
    {\textsuperscript{e}\footnotesize{Modified Julian date of the observation}}\\
    \multicolumn{12}{l}
    {\textsuperscript{f}\footnotesize{Right ascension in decimal degrees }}\\
    \multicolumn{12}{l}{\textsuperscript{g}\footnotesize{Declination in decimal degrees}}\\
    \multicolumn{12}{l}{\textsuperscript{h}\footnotesize{Measured spectroscopic redshift from DEIMOS observations}}\\
    \multicolumn{12}{l}{\textsuperscript{i}\footnotesize{Heliocentric-frame spectroscopic redshift}}\\
    \multicolumn{12}{l}{\textsuperscript{j}\footnotesize{Redshift quality flag (-2 = bad quality spectrum; 1 = no measured emission; 2 = uncertain redshift measurement; 3, 4 = good quality redshift measurement)}}\\
    \multicolumn{12}{l}{\textsuperscript{k}\footnotesize{F160W $H$-band magnitude (from {\it{HST}}/WFC3, \citealt{Finkelstein_2022})}}\\
    \multicolumn{12}{l}{\textsuperscript{l}\footnotesize{Median photometric redshift from \citet{Stefanon_2017}}}\\ \\
    \multicolumn{12}{l}{(This table is available in its entirety in a machine-readable form in the online journal. A portion is shown here for guidance regarding its form and content.)}\\
	\end{tabular}
\end{table*}

Observing conditions varied throughout the survey. In general, seeing ranged from roughly $0{\farcs}6$ to $1^{\prime\prime}$ with variable cloud cover. The DEIMOS detector is comprised of 8 CCDs, with each object spectrum spanning two chips (blue and red). One of the chips (CCD5) was inoperative during the 2020 observations and had an elevated level of read noise during the 2021 observations, resulting in decreased sensitivity (or a total loss of spectral coverage) at red wavelengths for approximately $\sim 25\%$ of the slits per mask. The number of targets per mask, for which the resulting spectra do not fall on CCD5 (i.e.~unaffected by this issue), are listed in Table~\ref{table:obs}.

Once the DEIMOS observations were completed, we reduced the entire dataset using the \texttt{spec2d}\footnote{\url{https://sites.uci.edu/spec2d/}} DEEP2/DEEP3 DEIMOS data reduction pipeline \citep{Newman_2013,Cooper_2012_spec2d}. Spectroscopic redshifts were measured using a custom template fitter that incorporates both an emission-line galaxy template (included to find redshifts for low-$z$ interlopers) as well as an asymmetric Gaussian profile to probe a single Ly$\alpha$ line where no other emission lines were detected. Examples of the best-fit templates for three high-$z$ Ly$\alpha$ emitters are shown in Figure \ref{fig:temp_ex}. The uncertainty reported in the redshift measurement is derived from the 1$\sigma$ error on the location of the peak Ly$\alpha$ emission determined from the fit. The quality of each redshift was visually inspected and given a quality code ($Q$) following the previous classification from the DEEP2/DEEP3 surveys. A quality code of $Q = -2$ indicates major detector/reduction issues, rendering at least half of the spectrum unusable. Out of the 276 total targets assigned $Q = -2$, 245 targets ($\sim 89 \%$) were placed on CCD5. Slits that had no detected emission or continuum are assigned a quality code of $Q = 1$. Upon visual inspection, targets with unclear or low-quality redshift measurements received a quality code of $Q = 2$. These objects would require follow-up analysis for redshift confirmation and are thus not reported in our final sample. Secure redshifts have a quality flag of either $Q = 3$ or $Q = 4$. Quality $Q = 4$ objects differ from $Q = 3$ by, upon visual inspection, having clear characteristics of an asymmetric Ly$\alpha$ profile (or multiple emission lines in the case of low-$z$ interlopers). 

\section{Redshift Catalog} \label{sec:data}

\begin{figure}
\centering
\includegraphics[width=0.9\linewidth]{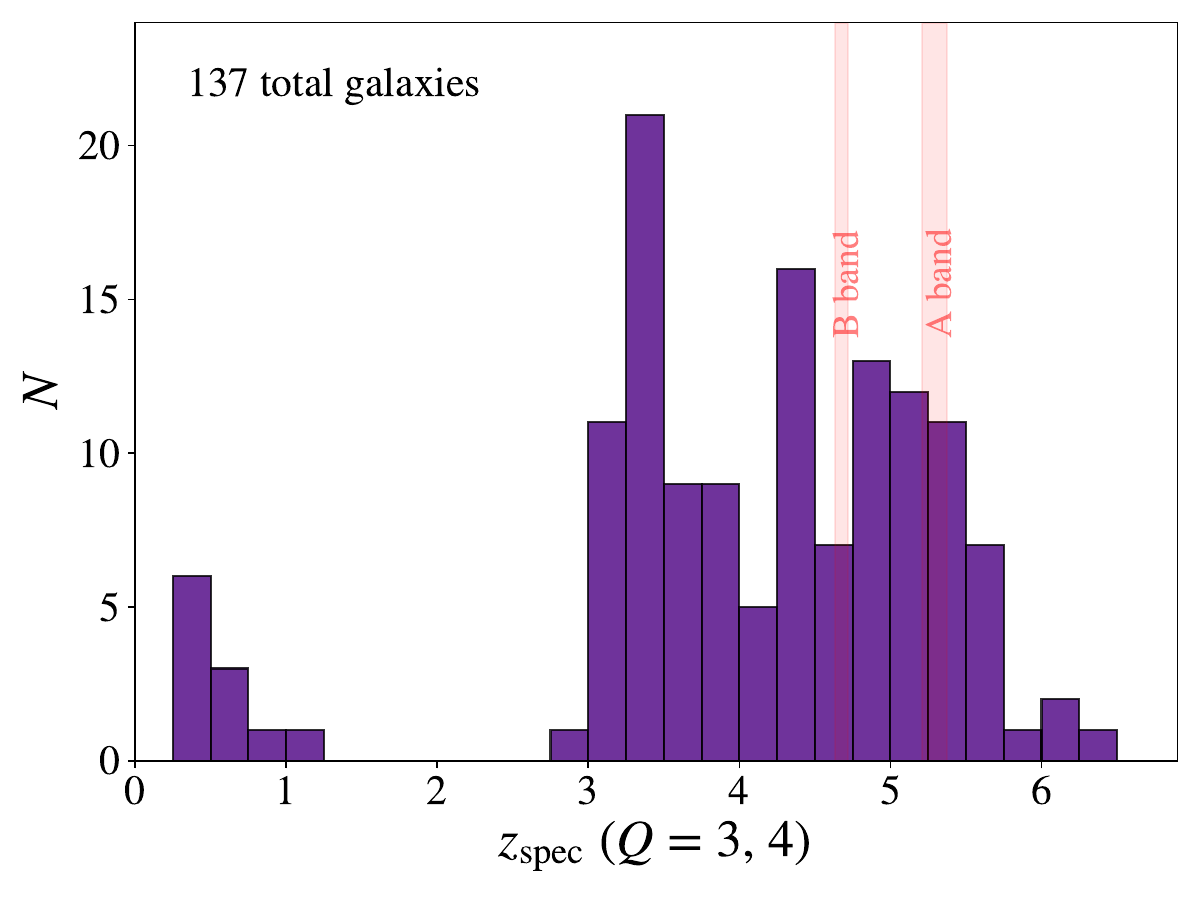}
\caption{Distribution of unique, heliocentric-corrected spectroscopic redshifts with secure quality flags ($Q = 3, 4$) for the targeted sample. Red vertical bands show the redshift ranges at which Ly$\alpha$ emission falls in the A or B atmospheric bands. Our survey yielded secure redshifts for a total of $137$ sources, including $126$ galaxies at $z > 2.8$.}
\label{fig:zhist} 
\end{figure}

\begin{figure*}
	\centering
    \includegraphics[width=0.33\linewidth]{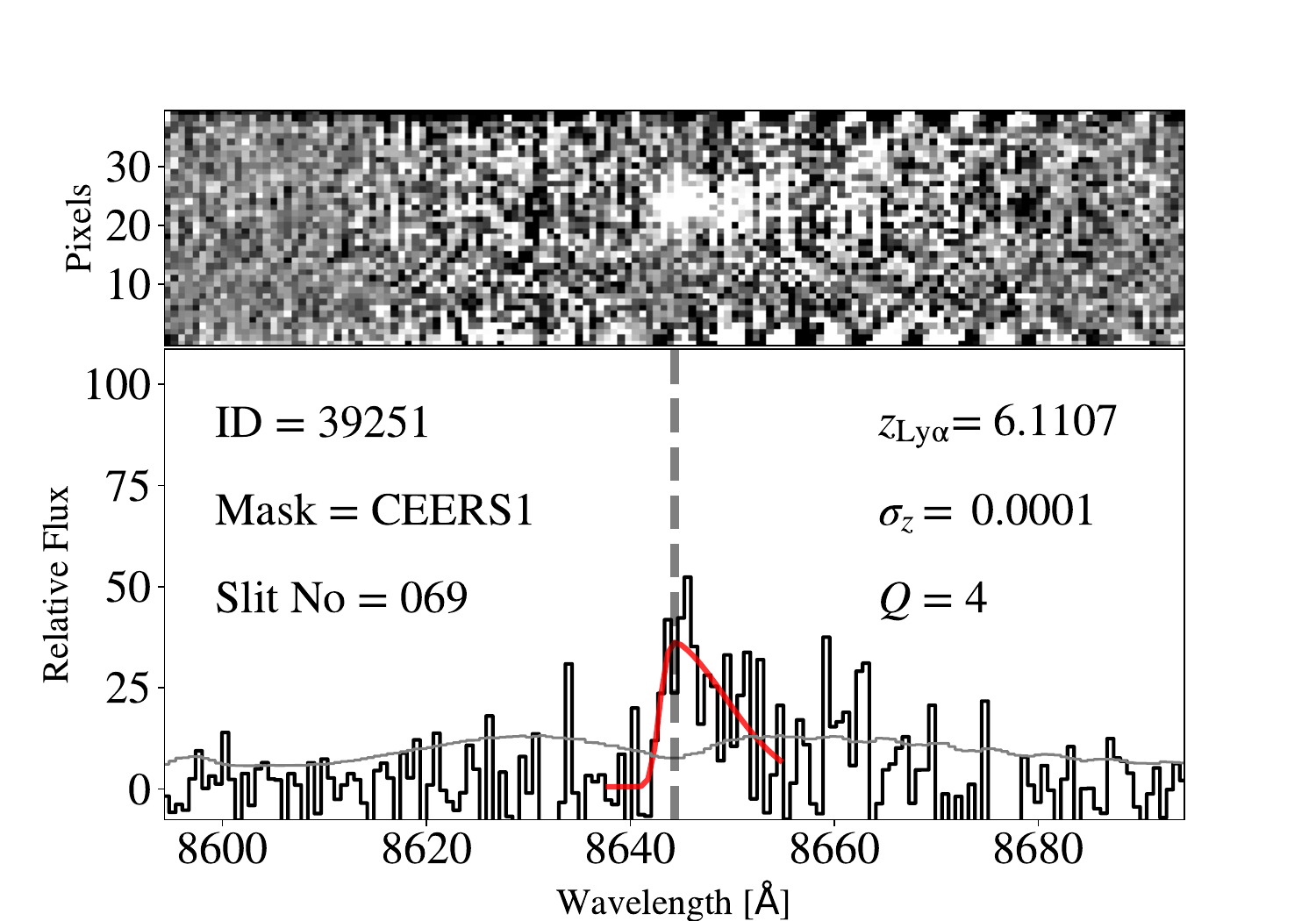}        \includegraphics[width=0.33\linewidth]{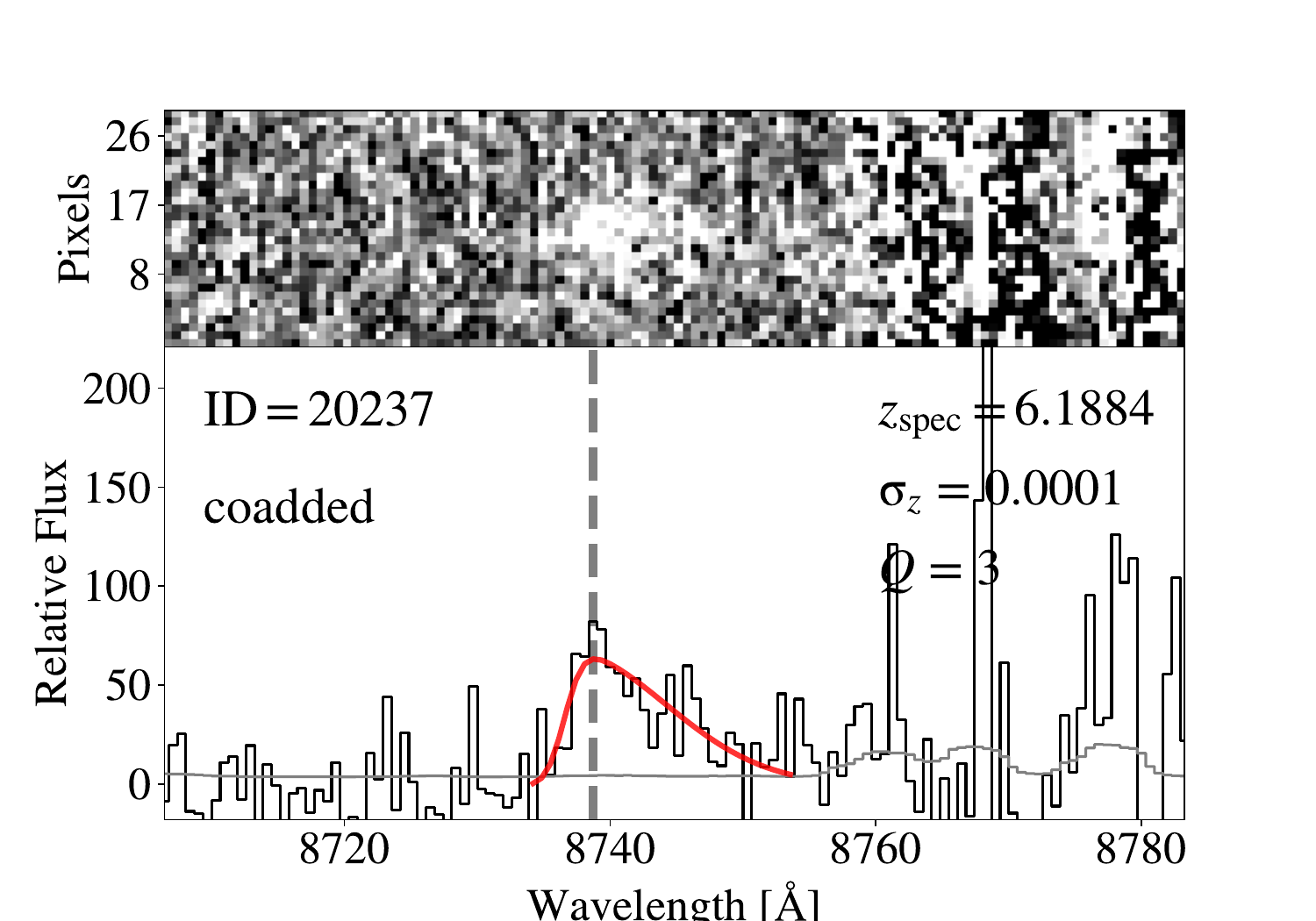}
    \includegraphics[width=0.33\linewidth]{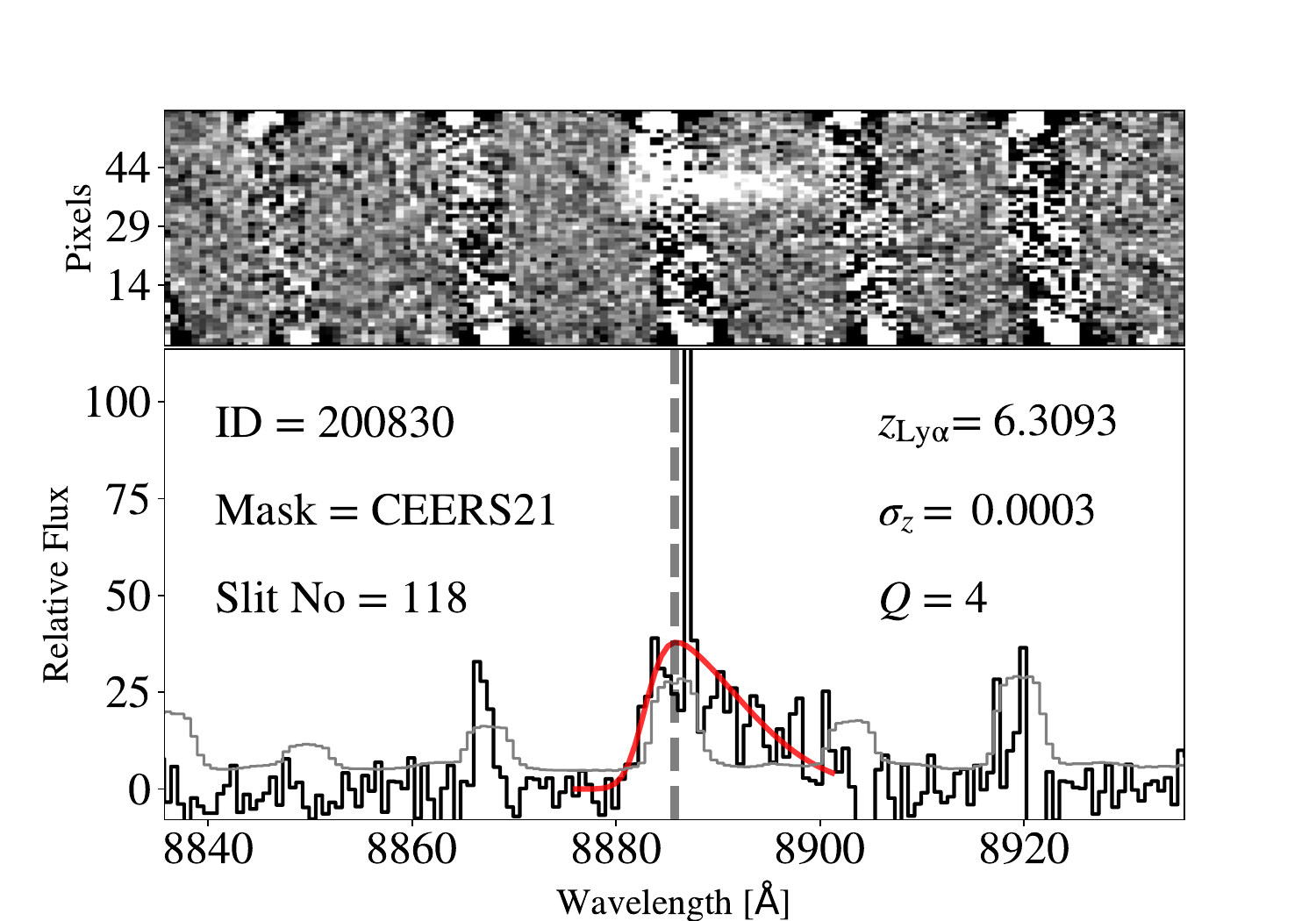}
    \includegraphics[width=0.33\linewidth]{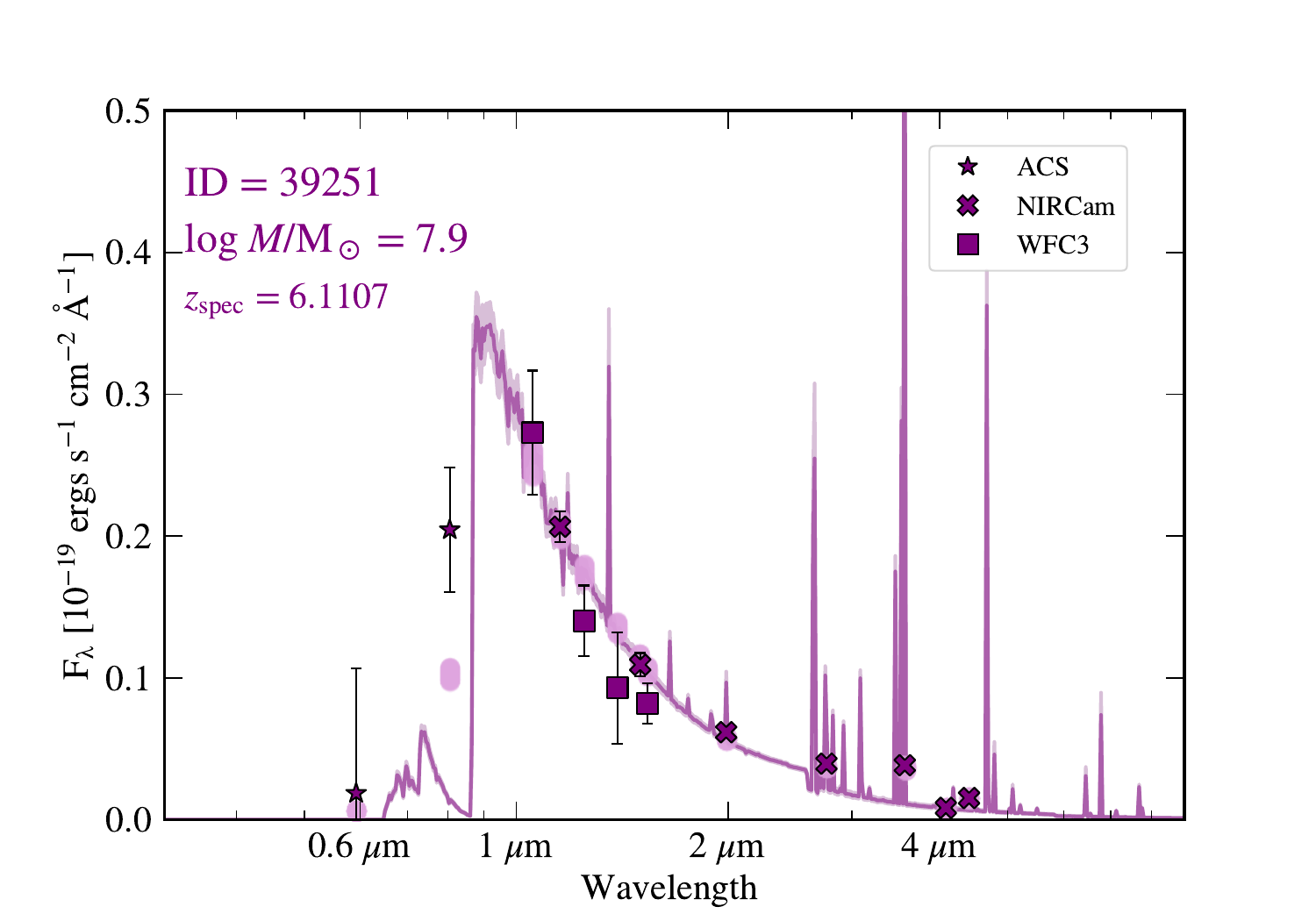}        \includegraphics[width=0.33\linewidth]{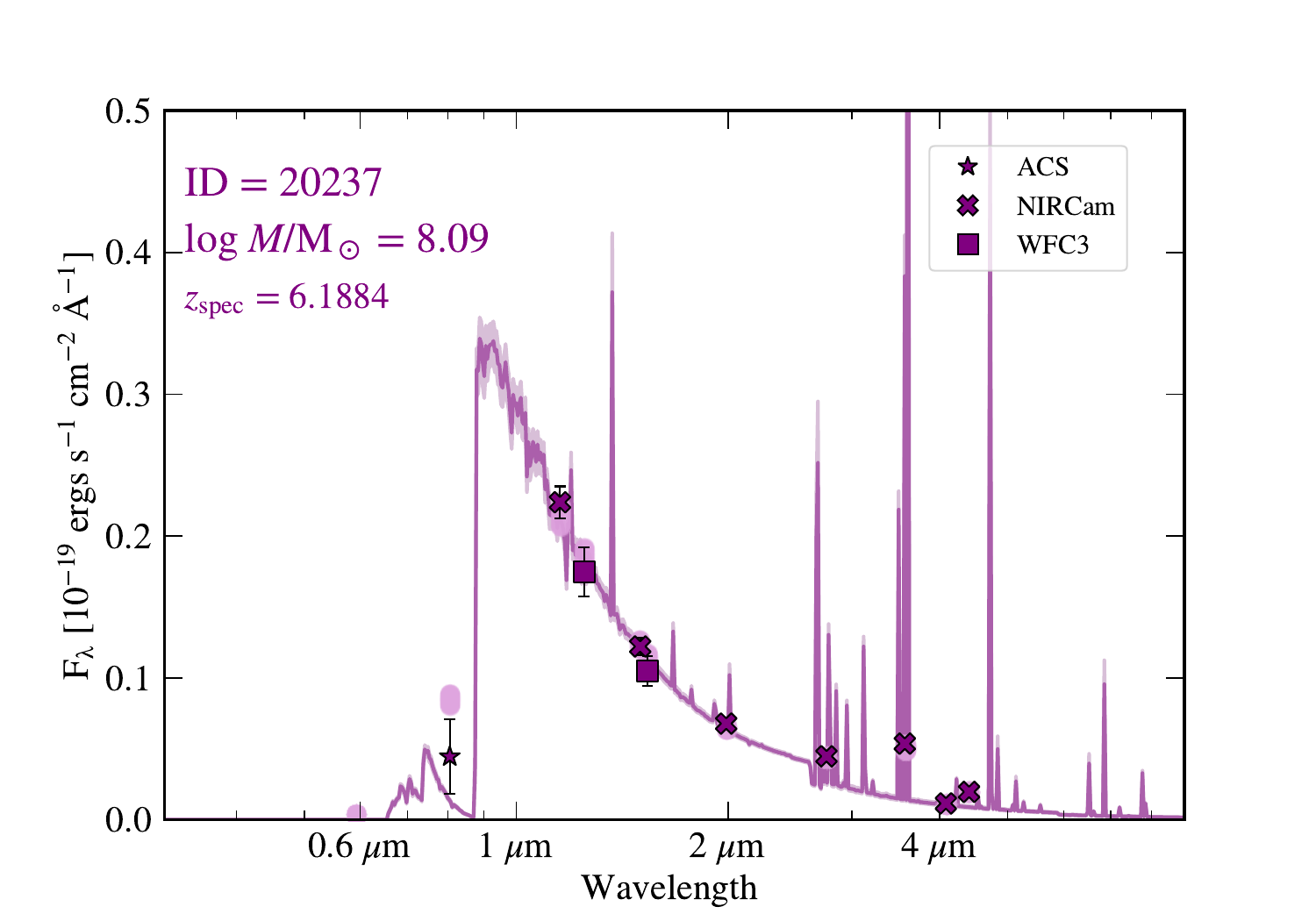}
    \includegraphics[width=0.33\linewidth]{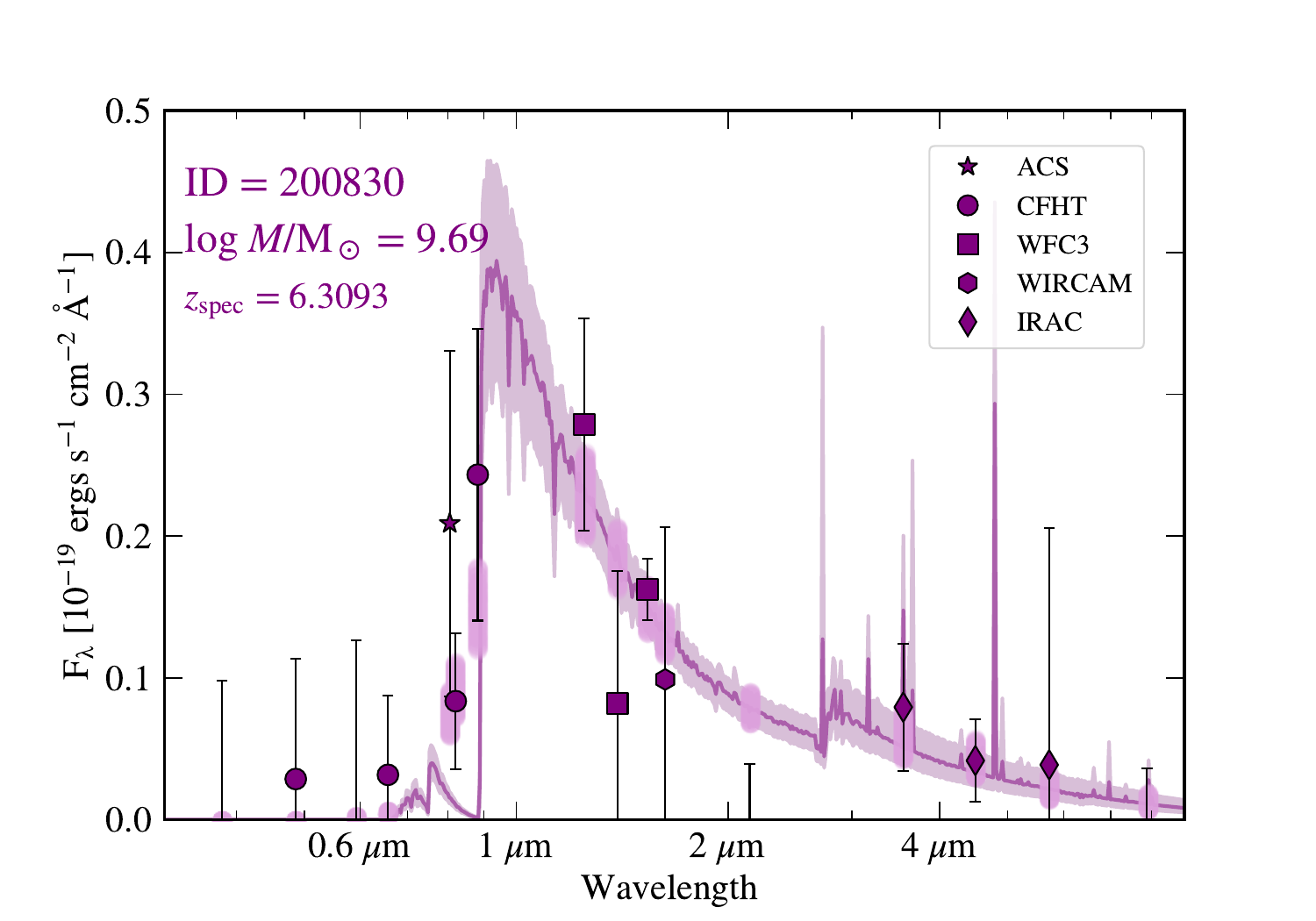}    
		\caption{Spectra and resulting SED fits for the three Ly$\alpha$ emitters at $z_\mathrm{spec} > 6$. Top panels show the observed 1D and 2D spectra, as in Figure \ref{fig:temp_ex}, along with the asymmetric Gaussian fit (shown in red) used to measure the redshift. The bottom panels show the photometric SED fits, at the measured $z_\mathrm{spec}$, from \texttt{Bagpipes}. For ID 39251 and 20237, we utilized the photometric data from the CEERS {\it{JWST}}/NIRCam imaging combined with {\it HST}/ACS and WFC3 \citep{Bagley_2023}. We use photometry from \citet{Stefanon_2017} for ID 200830 as it is located outside the CEERS footprint.
\label{fig:z6}}
\end{figure*}

We present the spectroscopic measurements from our Keck/DEIMOS observations in Table \ref{table: cat} (the full version is available on the electronic version of the Journal). In summary, from 947 unique targets we were able to secure a spectroscopic redshift of high quality ($Q = 3, 4$) for 137 galaxies. Of these, $126$ are Ly$\alpha$ emitters at $2.8 < z < 6.3$ (yielding a $13 \%$ success rate) with a mean redshift of $\overline{z} = 4.3$. Figure~\ref{fig:zhist} shows the full redshift distribution for objects with secure spectroscopic measurements. The sample includes 11 low-redshift galaxies ($z < 1.2$; all but one were originally targeted as high-$z$ candidates) along with three galaxies at $z > 6$ that probe the end of the Epoch of Reionization (EOR). 
In Figure \ref{fig:z6}, we show the 1D and 2D spectra of the three highest-redshift objects found in this survey along with photometry from CEERS \citep{Bagley_2023} and CANDELS \citep{Stefanon_2017}. We also include the best-fit SEDs set at each spectroscopic redshift using \texttt{Bagpipes} \citep{Canall_2018}. For one of the three objects, ID 20237, the spectroscopic redshift greatly changed the output SED, which previously was set to $z_\mathrm{phot} =1.436$ after misidentifying the Lyman break as the Balmer break (see further discussion of this object in \S \ref{subsec:compare}). The most recent version of our catalog, including the 1D and 2D spectra, can be downloaded directly from the survey webpage.\footnote{\url{https://sstawins.github.io/deeper_than_deep/}}

\subsection{Sources with Multiple Observations}\label{subsec:multiObs}

As mentioned in \S\ref{sec:selection}, $173$ targets were observed more than once. However, due to observing conditions, mask placement, and signal-to-noise from mask to mask, not all of these repeat observations resulted in multiple secure redshift measurements. Of the $173$ sources with multiple observations, only $35$ galaxies have two or more independent, secure ($Q = 3,4$) redshift measurements. 

The value of the redshift measured in repeated observations of a given galaxy can be affected by variation in the placement of the slit with respect to the galaxy as well as variation in the resulting S/N of the observed spectrum. To assess the uncertainty of our redshift measurements, we utilize the deviation in redshift for galaxies with multiple $z_{\rm spec}$ measurements. We limit this analysis to the $30$ sources with repeated observations yielding a secure redshift ($Q = 3,4$) inferred from Ly$\alpha$ emission (i.e.~excluding the $5$ low-$z$ sources with multiple $z_{\rm spec}$ measurements). We fit a normalized probability density function to the differences in the measured redshift ($\Delta z$), yielding a one-sided $1\sigma$ standard deviation of $0.0013$ ($\sim 389~{\rm km}~{\rm s}^{-1}$) and a two-sided standard deviation of $0.0007$ ($\sim 209 ~\rm km \ s^{-1}$). This uncertainty is 1-2 dex larger than the uncertainty associated with the fits to the observed emission line in a single observation, as  described in \S\ref{sec:obs}. In general, this analysis suggests that the redshifts (at $z > 2$) reported in our catalog are accurate to $\sim 1 \times 10^{-3}$ ($\sim 300~{\rm km}~{\rm s}^{-1}$). For low-$z$ sources ($z < 2$), where redshifts are largely determined by fits to multiple emission lines including [O{\scriptsize II}], H$\delta$, H$\beta$, [O{\scriptsize III}], and H$\alpha$, the typical redshift uncertainty is assumed to be $\sim60~{\rm km}~{\rm s}^{-1}$, similar to that of the DEEP3 survey which utilized the same instrument configuration and slit width (\citealt{Cooper_2011, Cooper_2012_deep3}, see also \citealt{Wirth_2004}).

In addition to enabling a study of spectroscopic redshift precision, repeated observations of sources within our survey allow for co-adding observations to produce higher signal-to-noise spectra. This is particularly interesting for sources that did not yield a secure redshift based upon any individual observation. One such source is 37653, which was targeted twice with Keck/DEIMOS, yielding a lower-quality redshift ($Q=1,2$) for each observation. To increase the signal-to-noise, we co-added the reduced, sky-subtracted 2D spectra, producing a combined observation with an effective exposure time of $7.7$~hours. We then extracted a 1D spectrum, using a boxcar extraction width of $\pm 3$ pixels ($\sim 0{\farcs}7$). Finally, we fit the resulting 1D spectrum, using the same asymmetric Gaussian fit described in \S\ref{sec:obs}, to find a best-fit, secure ($Q=3$) redshift of $z_{\rm spec} = 4.8998$. The resulting spectroscopic redshift is in excellent agreement with the photo-$z$ estimated from existing ground-based and {\it HST} imaging \citep[$z_{\rm phot} \sim 4.95$,][]{Stefanon_2017, Finkelstein_2022, Kodra_2023}. This source is also one of a small number of high-redshift galaxies -- with a confirmed spectroscopic redshift -- detected in the initial MIRI imaging for the CEERS survey \citep{Papovich_2023}. Recent NIRSpec prism observations (in December 2022) as part of the CEERS survey have confirmed our spectroscopic redshift (with $z_{\rm NIRSpec} = 4.89651$, Arrabal Haro et al.~in prep; see further discussion in \S\ref{subsec:compare}). By co-adding repeated observations for other sources in our target sample, we were able to measure secure redshifts for an additional $4$ galaxies (i.e.~$5$ in total, including IDs 37653, 30014, 27862, 20237, and 24687).

\subsection{Catalog Comparison to Literature} \label{subsec:compare}

\subsubsection{Connection to Photometric Catalogs}

\begin{figure}
	\centering
	\includegraphics[width=\linewidth]{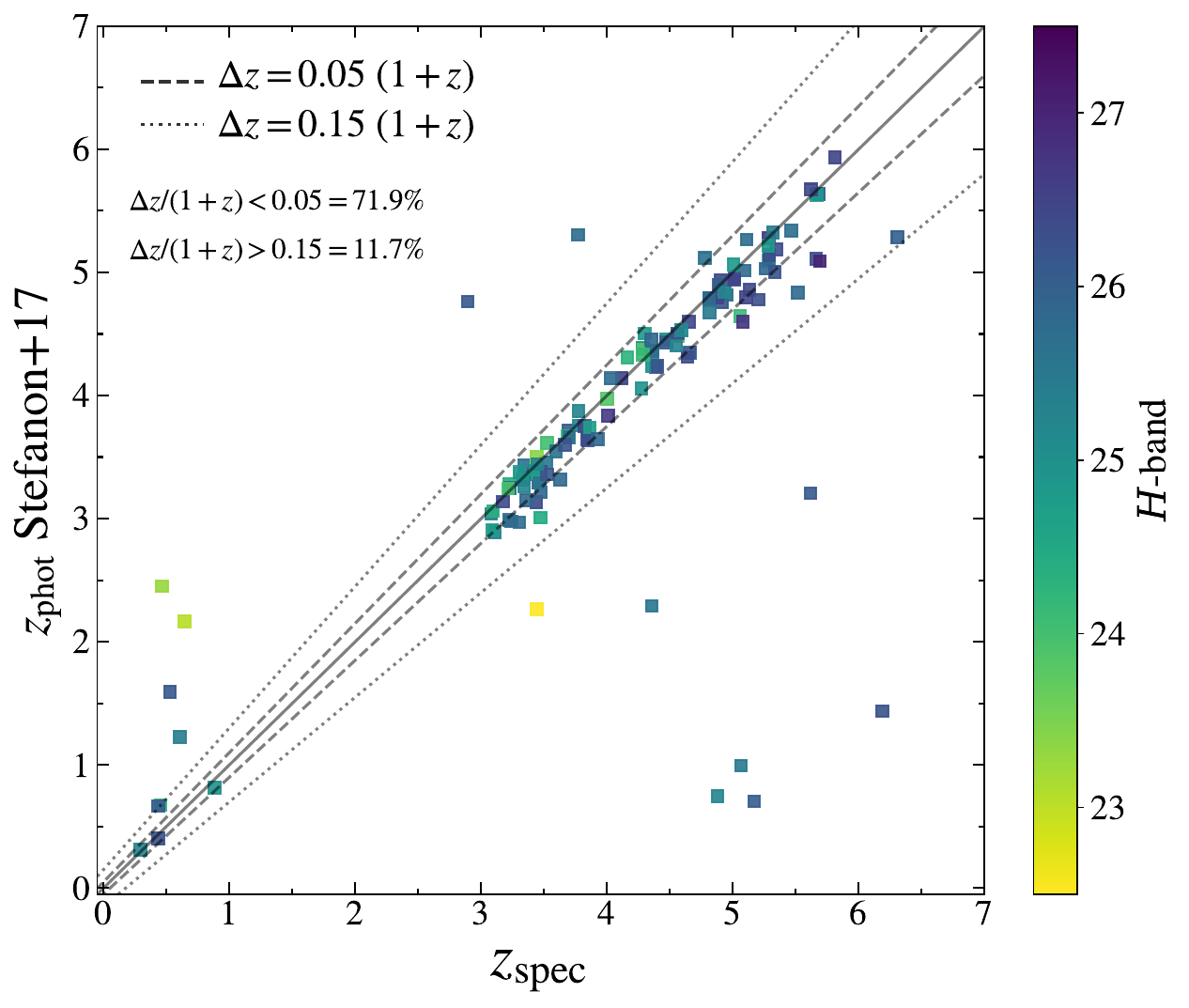}
	\includegraphics[width=\linewidth]{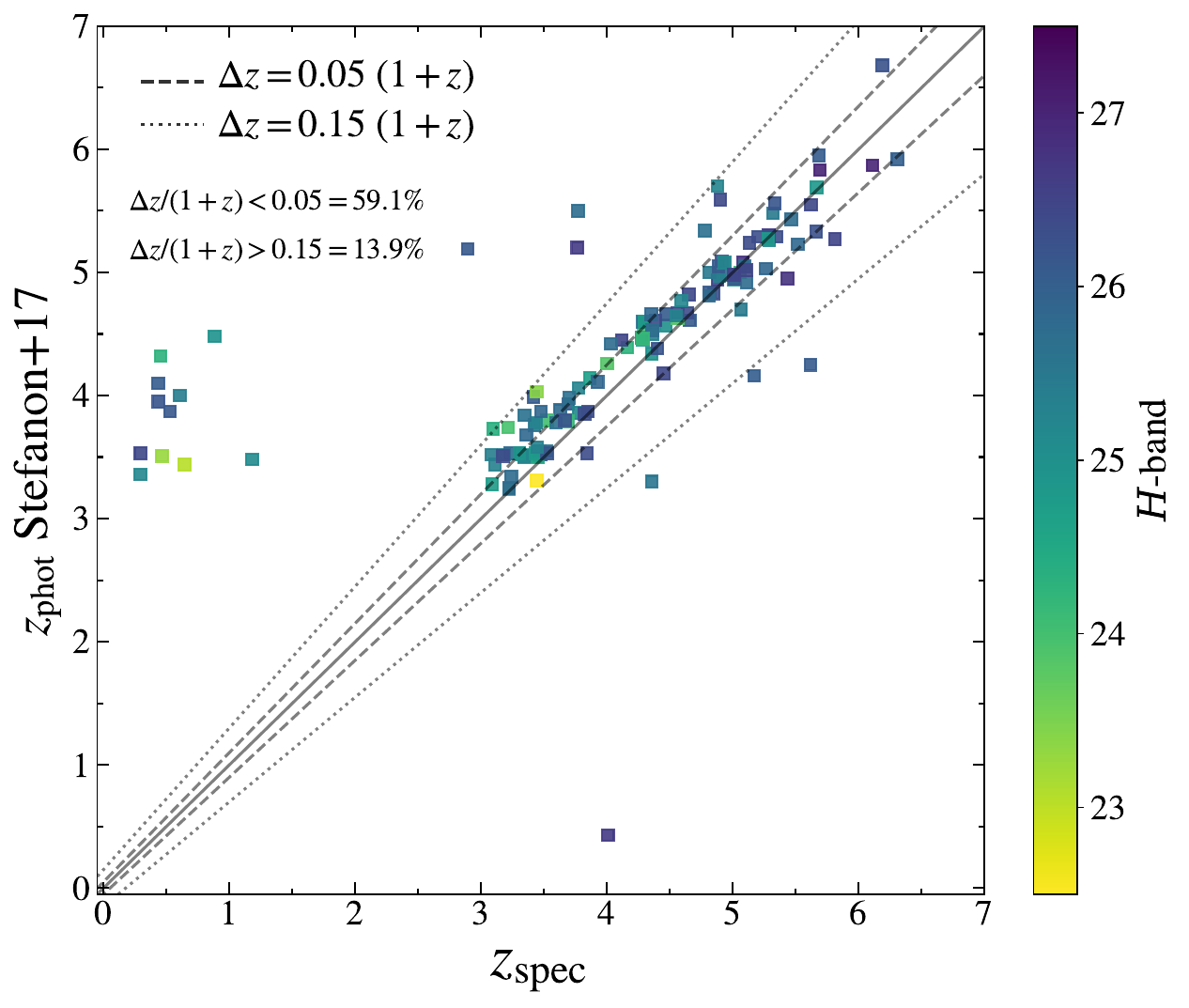}
	\caption{Median $z_\text{phot}$ for \citet{Stefanon_2017} (top) and \citet{Finkelstein_2022} (bottom) versus measured $z_\text{spec}$ from this work. Photometric redshfits from \citet{Finkelstein_2022} are the higher of the two dervied redshifts (with or without IRAC), reflecting our target sample (see \S \ref{sec:selection}). Dotted and dashed lines represent differences of $\Delta z > 0.15 (1+z)$ and $\Delta z > 0.05 (1+z)$, respectively. The color bar indicates the $H$-band magnitude ({\it{HST}}/WFC3 F160W) for each source. Overall, we find excellent agreement between the $z_\text{phot}$ and $z_\text{spec}$ measurements, with a small outlier fraction of $11.7\%$ and $13.9\%$ for \citet{Stefanon_2017} and \citet{Finkelstein_2022}, respectively. The median offset of the photometric redshifts in \citet{Stefanon_2017} from our spectroscopic redshifts is $\Delta z/(1+z) = 0.013$ with a $1\sigma$ standard deviation of $0.03$; similar to \citet{Finkelstein_2022} with $\Delta z/(1+z) = 0.02$ and $1\sigma$ standard deviation of $0.04$. \label{fig:zdiff}}
\end{figure}

To put our catalog into context with a large, photometric catalog in the EGS, we match our sample of spectroscopically confirmed galaxies to the \cite{Stefanon_2017} photometric catalog. We match the two catalogs by separation on the sky, requiring a maximum separation of $0{\farcs}2$. Out of 137 total galaxies with secure redshifts from this work, 130 are found in the \cite{Stefanon_2017} catalog, with the corresponding object ID provided in Table~\ref{table: cat}. We compare the expected median photometric redshift reported in \cite{Stefanon_2017} with the spectroscopic redshift from this work in Figure \ref{fig:zdiff}. For comparison, we also plot our spectroscopic redshifts versus the photometric redshifts with or without IRAC (see \S \ref{sec:selection}) from our target catalog \citep{Finkelstein_2022} in Figure \ref{fig:zdiff}. Overall, there is good agreement between our spectroscopic redshifts and the photometric redshifts from both catalogs. We find $71.9\%$ (92 galaxies) of photometric redshifts from \cite{Stefanon_2017} are within $\Delta z < 0.05 (1+z)$, and only $11.7\%$ (15 galaxies) exceed a maximum difference of $\Delta z > 0.15 (1+z)$. The median offset of these photometric redshifts from our spectroscopic measurements -- excluding significant outliers, $\Delta z > 0.15 (1+z)$ -- is $\Delta z/(1+z) = 0.013$ with a 1$\sigma$ standard deviation of $0.03$. This is similar to the results when comparing to the \cite{Finkelstein_2022} catalog, for which $59.1\%$ of the photometric redshifts are within $\Delta z < 0.05 (1+z)$ of the spectroscopic redshift, and $13.9\%$ are significant outliers outside $\Delta z > 0.15 (1+z)$. We find the median value of $\Delta z/(1+z)$ to be $0.02$ with a 1$\sigma$ standard deviation of $0.04$ excluding significant outliers.

In comparison to the photometric redshifts from both  \citet{Stefanon_2017} and  \citet{Finkelstein_2022}, there are a handful of outliers that have a $z_\mathrm{phot} < 1$ with a measured $z_\mathrm{spec} > 4$ (see Fig.~\ref{fig:zdiff}). These outliers are more prevalent with photometric redshifts from \citet{Stefanon_2017}, however this driven by the fact that photometric redshifts from \citet{Finkelstein_2022} shown in Fig.~\ref{fig:zdiff} are the higher of the two $z_\mathrm{phot}$ measurements, with or without IRAC, as used in target selection (see \S \ref{sec:selection}). The discrepancy in redshifts for the remaining outliers is primarily driven by SED modeling misidentifying the Lyman break as the Balmer break and/or by marginal detections at optical wavelengths. One such object is ID 20237, shown in Figure \ref{fig:z6}. This object has a median photometric redshift of $z_\mathrm{phot} = 1.436$ from \citet{Stefanon_2017}, which predicts a Balmer break at $\sim 8870$ \AA. However the spectroscopic redshift of this object is $z_\mathrm{spec} = 6.1884$, consistent with the Lyman break at a similar wavelength ($\sim 8700$ \AA).

\subsubsection{Synergy with other Spectroscopic Datasets}

Five galaxies with secure redshifts from this work (IDs 69719, 47173, 24711, 200415, 200814) correspond to sources with previously published spectroscopic redshifts. We find good agreement between our measured redshifts and the published values in all but one case. Two galaxies in our sample, ID 69719 at $z = 3.438$ and 47173 at $z = 3.304$, were also observed (in the near-IR) as part of the MOSDEF survey (ID$_\text{MOSDEF}$ 30847 and 13470, \citealt{Kriek_2015}), with measured redshifts of $z = 3.435$ and $z = 3.302$, respectively. One low-redshift emission line galaxy in this work (ID 24711, $z = 0.2948$) was also included in DEEP2 Data Release 4 (DR4, ID$_\text{DR4}$ 12028700; \citealt{Newman_2013}) with $z = 0.2955$. Finally, two objects -- ID 200415 at $z = 0.4639$ and ID 200814 at $z = 0.6426$ -- were included in the spectroscopic catalog from the 3D-HST survey (ID$_\text{3D-HST}$ 21636 and 29958, \citealt{Momcheva_2016}). Our measured redshift for the latter source (ID 200814) is in good agreement with the measurement of $z = 0.6738$ from the lower-resolution 3D-HST grism spectrum. For ID 200415, however, we find a significant difference in redshift, relative to the 3D-HST measurement of $z = 0.73945$. The WFC3 G141 grism spectrum for this object includes an emission feature towards the blue (near $\sim 1.14~\mu{\rm m}$) that is identified as H$\alpha$ (yielding $z = 0.73945$). This portion of the 3D-HST grism spectrum, however, suffers from a high level of contamination associated with a nearby bright source, such that the emission feature identified as H$\alpha$ is likely spurious. In our Keck/DEIMOS spectrum, we detect a multitude of emission features, including [O{\scriptsize II}], H$\beta$, and [O{\scriptsize III}] --- \emph{as well as} H$\alpha$ and [N{\scriptsize II}]. 

With the exception of ID 200415, the difference in the measured spectroscopic redshift between that of our survey and previously published values for these 4 sources ranges from $\Delta z = 0.0007-0.03$. These small differences in the measured $z$ are consistent with the differences in spectral resolution, rest-frame spectral features sampled, and potential variation in slit placement between our observations and those of the MOSDEF, DEEP2, and 3D-HST surveys.    

During December 2022, the CEERS team observed the EGS with NIRSpec. The CEERS NIRSpec multi-object slit masks were primarily designed around promising $z \gtrsim 8$ candidates observed with previous imaging campaigns from the first set of CEERS imaging. These observations covered $1-5 \mu$m wavelengths using both medium resolution gratings ($R = 1000$; G140M/F100LP, G235M/F170LP, and G395M/F290LP) and the prism ($R = 30-300$) for a total integration time of $3107$~s. While the primary goal for these observations were to find $z \gtrsim 8$ galaxies, some of the targeted objects were instead confirmed to have lower spectroscopic redshifts ($z_\mathrm{spec} = 4-6$) that lie within the redshift range of our catalog. The NIRSpec observations at this redshift range covers a suite of diagnostic spectral features, such as the $4000$\AA-break, [O{\scriptsize II}], and H$\alpha$, which can be used in combination with our observations to measure the correlation with Ly$\alpha$ emission and important galaxy conditions. Two galaxies with a measured spectrum (ID 10496 and 14628 from \citealt{Stefanon_2017}) are also found in our catalog (ID 47173 and 37653, respectively). ID 47173 was observed twice in 2020 and 2021 with DEIMOS, resulting in two redshift measurements ($z_\text{spec} = 3.3038$ with $Q=3$ and $z_\text{spec} = 3.3056$ with $Q=2$). NIRSpec spectroscopy is consistent with the $Q=3$ measured redshift by $\Delta z = 0.002$, with $z_\text{NIRSpec} = 3.30186$ (MSA ID 11699, Arrabal Haro et al.~in prep). The difference in redshift in part represents the offset between the Ly$\alpha$ emission ($\Delta v_\mathrm{Ly\alpha}$) and the associated metal lines observed in NIRSpec spectroscopy, which better trace the systemic redshift of the galaxy. In this case, the observed offset in Ly$\alpha$ emission for this galaxy is $580$ km s$^{-1}$. This is within the range of $\Delta v_\mathrm{Ly\alpha}$ found by \citet{Marchi_2019} around $z \sim 3.5$ star-forming galaxies, with offsets up to $800$ km s$^{-1}$ and an average offset of $\langle \Delta v_\mathrm{Ly\alpha} \rangle = 358$ km s$^{-1}$. 

For ID 37653, both DEIMOS observations yielded low-quality ($Q = 2$) redshift measurements. As discussed previously in \S\ref{subsec:multiObs}, we use the co-added spectrum of two individual DEIMOS observations to measure a secure redshift for this galaxy. We find the spectroscopic redshift to be $z_\mathrm{spec} = 4.8998$, which agrees with the NIRSpec observations at $z_\mathrm{NIRSpec} = 4.89651$ (MSA ID 707, Arrabal Haro et al.~in prep). This galaxy also lies close ($\sim 1-2$ comoving Mpc, in projection) to a recently identified overdense region at $4.5 < z < 5.5$ in the EGS \citep{ArrabalHaro2023,ArrabalHaro2023b,Naidu2022}. We find 7 other galaxies as within this region, all with $4.8 < z_\text{spec} < 5.0$. Studies of high-redshift proto-clusters using the Millennium cosmological simulations show that such clusters at $z \sim 5$ could be as large as $5-20$ cMpc \citep{Chiang_2013,Muldrew_2015}. Taken together, the redshift and the location of 37653 could indicate that this galaxy is a member of this overdensity at $z \sim 5$. 

Using the redshift given by NIRSpec for 37653, we find the offset of Ly$\alpha$ emission is large ($\langle \Delta v_\mathrm{Ly\alpha} \rangle =$ 970 km s$^{-1}$) compared to offsets found by \citet{Cassata_2020} around Ly$\alpha$ emitting galaxies at $z = 4.4-5.7$ ($\langle \Delta v_\mathrm{Ly\alpha} \rangle = 377$ km s$^{-1}$ with a scatter of $329$ km s$^{-1}$). In that work, none of the Ly$\alpha$ offsets exceed $\sim$800 km s$^{-1}$ for $z = 4.4-5.7$. More work is needed to identify the cause of this high-velocity offset.

\subsection{Redshift Success}

\begin{figure*}
	\centering
		\includegraphics[width=\linewidth]{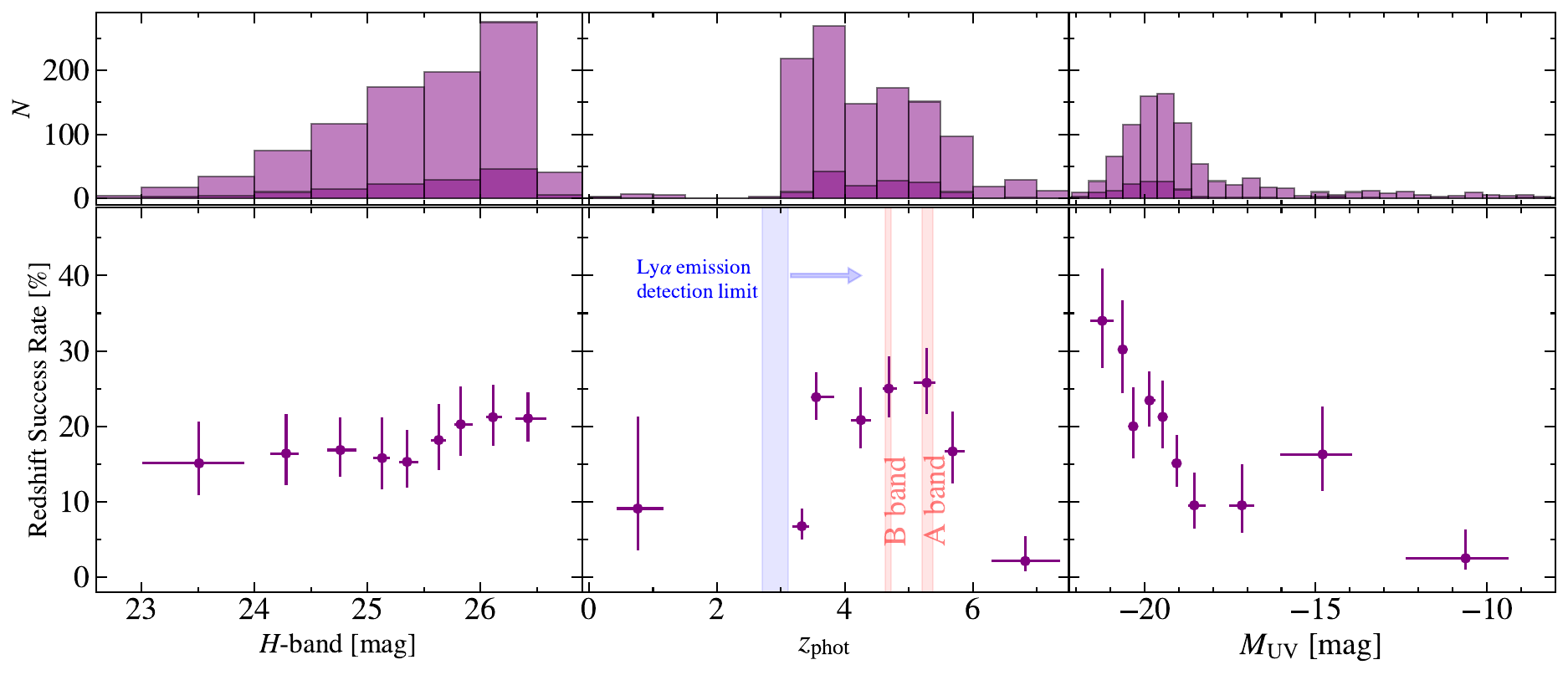}
		\caption{Redshift success rate (number of $Q = 3,4$ sources compared to targets observed) as a function of $H$-band apparent magnitude (left), photometric redshift (center), and UV absolute magnitude (right). Light purple histograms above display the number of sources with a given $H$-band, $z_\text{phot}$, and $M_\text{UV}$, with equally spaced bins of $0.5$ for all measurements. The corresponding dark purple histograms illustrate the number of $Q = 3,4$ sources in each bin. The success rate is calculated using variable bin sizes, adopting a minimum size of 50 total targeted sources with the exception of the lowest-$z$ bin. The $x$ error bars designate the $20^{\rm th}$ and $80^{\rm th}$ percentiles of the distribution within each derived bin. The $y$ error bars represent the 1$\sigma$ Wilson score interval (uncertainty; \citealt{Wilson_1927}) of a binomial distribution. The minimum effective Ly$\alpha$ detection limit, calculated using the bluest wavelengths probed by a typical DEIMOS spectrum ($\sim4500$-$5000$ \AA), is shown as the blue shaded region and corresponding arrow.   
\label{fig:zsuccess}}
\end{figure*}

Lastly, in this section, we discuss the success rate of our observations and compare those to preliminary selection criteria and other physical parameters. In Figure~\ref{fig:zsuccess}, we present the redshift success rate for our survey as a function of $H$-band apparent magnitude, $z_\text{phot}$, and absolute UV magnitude; where the redshift success rate is defined to be the number of sources with a secure ($Q =3,4$) redshift divided by the total number of targets observed. For this analysis, we exclude objects with $Q=-2$ as they are effectively unobserved.  

We first calculate the redshift success rate as a function of $H$-band magnitude and photo-$z$ from our target sample \citep{Finkelstein_2022}. In general, the redshift success rate is $\sim 15\%$ for targets with $H \lesssim 25$. Although this slightly increases up to $\sim 20 \%$ for dimmer targets, and with consideration of the 1 $\sigma$ uncertainty for each bin, we find no significant dependence of redshift success on apparent $H$-band magnitude. This implies apparent $H$-band magnitudes are not a biased tracer of Ly$\alpha$ emission at this redshift range. 

The redshift success does vary with $z_\text{phot}$ across our targeted redshift range, mostly due to observational constraints. Near the effective low-$z$ detection limit for Ly$\alpha$ emission given our DEIMOS setup ($z \sim 3.1$), the success rate reaches as low as $6.8\%$. Meanwhile, at $z \gtrsim 5.5$, we find a decrease in the $z$ success rate, dropping to $2.1\%$ for $z_\text{phot} > 6$, likely due to the increase in sky emission at redder wavelengths. Low-redshift sources ($z < 2$) have a low success rate ($9.1 \%$), however the uncertainty is larger due to the smaller number of total targets. 

The average redshift success rate is relatively flat from $3.5 < z_\text{phot} < 5.5$. For this redshift range, the average success rate is $24.1\%$ and the 1$\sigma$ standard deviation of the distribution is $\sim 1.5 \ \%$. For $z_\text{phot} > 3$, we are exclusively probing Ly$\alpha$ emitters, hence the redshift success rate could be \textit{related} to a Ly$\alpha$ detection fraction ($f_\mathrm{Ly\alpha}$) as measured in previous works \citep{Stark_2010,Stark_2011,Ono_2012}. However, here we are not taking into account the variability in instrument sensitivity as a function of wavelength, which would directly translate into a corresponding variation in the sensitivity to Ly$\alpha$ detection as a function of $z$. Therefore, more work must be done to make a direct comparison between our measured redshift success rate and existing measurements of the Ly$\alpha$ detection fraction. Overall, this analysis shows our survey had excellent success around our desired redshift range. 

Using the \citet{Stefanon_2017} photometric catalog, we can also analyze the redshift success rate as a function of other physical parameters such as stellar mass, SFR, rest-frame $U-V$ color, and absolute UV magnitude ($M_\text{UV}$). We find no significant dependence of the redshift success rate on stellar mass and SFR. Conversely, the redshift success rate does strongly depend on rest-frame $U-V$ color and absolute UV magnitude. For rest-frame $U-V$ color, we find that the success rate increases towards bluer colors, reaching $38\%$ for $U-V \sim -0.1$ and decreasing to $\sim 5\%$ at $U-V \gtrsim 1.2$. As shown in the right-most panel of Figure~\ref{fig:zsuccess}, we also find that the redshift success rate increases towards brighter $M_\text{UV}$ magnitudes, reaching as high as $34\%$ for $M_\text{UV} \sim -21$. Although previous work by \citet{Stark_2010} found that the Ly$\alpha$ fraction from $z = 3-6$ to be higher at fainter $M_\text{UV}$, we again caution our redshift success rate is not directly comparable to a Ly$\alpha$ detection fraction. In their previous work, \citet{Stark_2010} calculated the Ly$\alpha$ detection fraction for sources with equivalent widths (EW) $>50$\AA. While work to measure Ly$\alpha$ EWs for our sample is ongoing, preliminary measurements show we are probing Ly$\alpha$ emitters below $50$~\AA\ for $M_\text{UV}$ magnitudes brighter than $-18$.

\section{Conclusion} \label{sec:conc}
In this work, we targeted 947 high-redshift galaxies ($z_\text{phot} > 3$) in the EGS with Keck/DEIMOS. In total, we measured spectroscopic redshifts for 137 galaxies, including 126 confirmed Ly$\alpha$ emitters at $2.8 < z < 6.3$. This catalog significantly expands the number of spectroscopically confirmed galaxies in the EGS field at $z_\text{spec} > 3$.

Overall, we find good agreement with our spectroscopic redshifts and photometric catalogs in the literature \citep{Finkelstein_2022,Stefanon_2017}. A comparison between redshifts yields a small difference ($\Delta z / (1+z) < 0.05$) of 59.1$\%$ and 71.9$\%$, respectively. We also find 4 galaxies that have spectroscopic redshifts from other surveys in literature (i.e. MOSDEF, 3D-HST, and DEEP2), with the difference in spectroscopic redshifts ranging from $\Delta z = 0.0007 - 0.03$.

This work comes at an opportune time, given the recently completed observations from the {\textit{JWST}} ERS program CEERS. With the influx of photometric data from {\textit{JWST}}, it becomes increasingly useful to have spectroscopic constraints to constrain photometric SED fits. Furthermore, spectroscopic redshifts are more reliable than those measured with photometric imaging, allowing improved target selection for future observations. 

In December 2022 CEERS observed high-redshift galaxies, detecting faint emission from galaxies out to $z = 4-6$ using the NIRSpec multi-object spectrograph. Two galaxies targeted during these NIRSpec observations were also found in this catalog, with agreeing redshifts. Emission lines from these {\textit{JWST}} observations will allow for analysis of gas conditions in galaxies at much higher redshift than previously studied. Together with our DEIMOS observations, we can start to measure the correlation between the dependence on Ly$\alpha$ emission with important galaxy conditions at $z > 4$. As demonstrated in \S\ref{subsec:compare}, these measurements together could also lead to further characterization of Ly$\alpha$ velocity offsets for galaxies at $z = 3-6$. 

Given the importance of spectroscopic measurements of galaxies at $z > 4$ in the EGS field, we are currently working on additional ground-based observations in collaboration with the Web Epoch of Reionization Lyman-alpha Survey (WERLS; PI: C. Casey and J. Kartaltepe). The ongoing program, which was allocated time in 2022A/B and 2023A/B, is targeting the EGS Field with Keck/LRIS and Keck/MOSFIRE with the goal of detecting Ly$\alpha$ emitters at the latter half of the EOR ($5.5 < z < 8$; see the preliminary catalogs from \citealt{cooper_2023} for the Keck/MOSFIRE observations and Urbano Stawinski et al. in prep for the Keck/LRIS observations). This future work will expand upon the efforts of this paper and significantly add to known galaxies at $z > 4$ in the EGS.

\section*{Acknowledgements}

The authors wish to recognize and acknowledge the very significant cultural role and reverence that the summit of Maunakea has always had within the indigenous Hawaiian community. We are most fortunate to have the opportunity to conduct observations from this mountain. MCC and SMUS acknowledge support from the National Science Foundation through grant AST-1815475. PGP-G acknowledges support from Spanish Ministerio de Ciencia e Innovaci\'{o}n MCIN/AEI/10.13039/501100011033 through grant PGC2018-093499-B-I00.

\section*{Data Availability}

The data underlying this article are available in the survey GitHub webpage: \url{https://sstawins.github.io/deeper_than_deep/}.

\bibliographystyle{mnras}
\bibliography{biblio}

\bsp	
\label{lastpage}
\end{document}